\documentclass[twocolumn,trackchanges]{aastex63}

\newcommand{\mum}{{\mu\mathrm{m}}}

\usepackage{hyperref}
\usepackage{txfonts}
\usepackage{booktabs}
\usepackage{float}
\usepackage{color} 

\iftrue

\definecolor{comment}{rgb}{0,0,1}

\fi

\newcommand{\ProgramOne}{077.C-0662}
\newcommand{\ProgramTwo}{079.C-0101}
\newcommand{\ProgramThree}{081.C-0020}

\newcommand{\Gaia}{\emph{Gaia}}
\newcommand{\TESS}{\emph{TESS}}

\newcommand{\Konkoly}{Konkoly Observatory, Research Centre for Astronomy and Earth Sciences, E\"otv\"os Lor\'and Research Network (ELKH), Konkoly-Thege Mikl\'os \'ut 15-17, 1121 Budapest, Hungary}

\newcommand{\MPIA}{Max Planck Institute for Astronomy, K\"onigstuhl 17, D-69117 Heidelberg, Germany}

\newcommand{\ELTE}{ELTE E\"otv\"os Lor\'and University, Institute of Physics, P\'azm\'any P\'eter s\'et\'any 1/A, 1117 Budapest, Hungary}
\newcommand{\Leiden}{Leiden Observatory, Leiden University, P.O. Box 9513, 2300 RA Leiden, The Netherlands}

\shorttitle{An infrared interferometric study of DI~Cha~A}
\shortauthors{Juh\'asz et al.}
\graphicspath{{./}{figures/}}

\begin{document}


\title{A gap at 1 au in the disk of DI Cha A revealed by infrared interferometry
\footnote{
    Based on observations collected at the European Organisation for Astronomical Research in the Southern Hemisphere under ESO programs {\ProgramOne}, {\ProgramTwo}, and {\ProgramThree}.
  }
}

\author[0000-0003-1719-8503]{T\'\i{}mea Juh\'asz}
\affiliation{\Konkoly}
\affiliation{\ELTE}
\email{juhasz.timea@csfk.org}

\author[0000-0001-6015-646X]{P\'eter \'Abrah\'am}
\affiliation{\Konkoly}
\affiliation{\ELTE}

\author{Attila Mo\'or}
\affiliation{\Konkoly}

\author[0000-0003-2835-1729]{Lei Chen}
\affiliation{\Konkoly}

\author[0000-0001-7157-6275]{\'Agnes K\'osp\'al}
\affiliation{\Konkoly}
\affiliation{\ELTE}
\affiliation{\MPIA}

\author[0000-0003-4989-575X]{J\'ozsef Varga}
\affiliation{\Leiden}
\affiliation{\Konkoly}

\author[0000-0001-5573-8190]{Zsolt Reg\'aly}
\affiliation{\Konkoly}

\author[0000-0002-4612-5824]{Gabriella Zsidi}
\affiliation{\Konkoly}
\affiliation{\ELTE}

\author[0000-0001-5449-2467]{András P\'al}
\affiliation{\Konkoly}

\begin{abstract}
DI Cha A is K0-type pre-main sequence star, the brightest component of a quadruple stellar system. Here we report on a detailed study of this star based on archival VLTI/MIDI and VLTI/PIONIER infrared interferometric observations, as well as optical--infrared photometric monitoring from ground-based and space-born instruments. We determined the structure of the circumstellar disk by fitting simultaneously the interferometric visibilities and the spectral energy distribution, using both analytical models and the radiative transfer code RADMC-3D. The modeling revealed that the radial density distribution of the disk appears to have a gap between 0.21 and 3.0 au. The inner ring, whose inner size coincides with the sublimation radius, is devoid of small, submicrometer-sized dust grains. The inner edge of the outer disk features a puffed-up rim, typically seen in intermediate-mass stars. Grain growth, although less progressed, was also detected in the outer disk. The inner ring is variable at mid-infrared wavelengths on both daily and annual timescales, while the star stays remarkably constant in the optical, pointing to geometrical or accretion changes in the disk as possible explanation for the flux variations.
\end{abstract}


\section{Introduction} \label{sec:Introduction}

According to the current paradigm, both the low-mass ($<$2~M$_{\odot}$) T~Tauri and the intermediate-mass  (2$<$M$<$8~M$_{\odot}$) Herbig Ae/Be stars are surrounded by protoplanetary disks during the first few million years of their evolution. In the class II phase the central star is already well visible at optical wavelengths, while its disk emits in the thermal infrared. In the simplest case, the disk has a continuous density distribution from the dust sublimation radius out to several hundreds of au, radiating in the whole infrared regime. Infrared spectra from the Spitzer Space Telescope, however, revealed disks where the spectral energy distribution (SED) shows unexpectedly low mid-infrared emission, interpreted as signature of an inner cavity in the disk, probably marking disk dispersal at a late evolutionary stage  \citep[transitional disks, e.g.,][]{williamscieza2011}. In some of these cases, strong near-infrared emission was still detected, leading to the concept of pre-transitional disks, when an inner disk or ring surrounds the young star within a few au distance, separated from an extended outer disk by a wide gap \citep{2010ApJ...717..441E}. 

In the ALMA era, the presence of circular narrow gaps in protoplanetary disks, typically on a few tens of au scale, turned out to be ubiquitous. Well-known examples are HL~Tau \citep{2015ApJ...808L...3A} and TW~Hya \citep{2016ApJ...819L...7N}. It is still an open issue whether the gaps are related to forming planets, or caused by hydrodynamic or other effects. Gaps or inner holes at few au scales, in the zone of terrestrial planet formation, remained  largely unresolved for ALMA, but could be reached via infrared interferometry. \citet{2015A&A...581A.107M} and \citet{2018A&A...617A..83V} analysed archival VLTI/MIDI observations for a large sample of T Tauri and Herbig Ae/Be stars, and suggested the existence of a disk population exhibiting au-scale cavities or gaps. New data obtained with the second generation interferometer VLTI/MATISSE are confirming the complex inner structure of protoplanetary disks, including even azimuthal asymmetric structures, like vortices \citep[e.g.,][]{2021A&A...647A..56V}.

Pre-main sequence stars usually show characteristic temporal changes in their luminosity. Their SED is  the combined emission of the star and the disk. The possible variability in brightness of the young central star not only causes fluctuations in the optical emission, but the change in radiation also affects the disk, so the variability is observable at infrared wavelengths as well. The silicate feature around 10$\,\mu$m provides further important information about the system: the amplitude and shape of the emission feature in this wavelength range depends on the material and size of the dust particles.

DI Chamaeleontis (DI Cha) is a good candidate for a complex inner disk structure. It is a quadruple system \citep{2013A&A...557A..80S} whose  primary star, DI Cha A, has a spectral type of K0 and a mass of $2.08\,M_\odot$  \citep{2017A&A...604A.127M}, and its emission dominates the SED at all wavelengths. A companion object was observed at a separation of $4\farcs6$ \citep{1993A&A...278...81R}, and later it was revealed that the companion itself is a double star \citep{2008ApJ...683..844L}. They are marked as the B and C components, orbiting each other at a separation of  $0\farcs06$. \cite{2013A&A...557A..80S} estimated that both B and C are dwarf stars with M$5.5\pm1.5$ spectral types. The most recently discovered star in the system, marked as D, is a close companion of DI~Cha~A, detected at a separation of approximately $0\farcs2$. DI Cha D is an M$6\pm1.5$ spectral type star and its estimated mass is $ \geq 67\ M_{Jup}$ \citep{2013A&A...557A..80S}.

The parallax of DI~Cha (A+D) in \Gaia~eDR3 is $5.291\pm0.013$\,mas, corresponding to a distance of $188.65^{+0.45}_{-0.48}$\,pc \citep{2021AJ....161..147B}. The \Gaia~measurement can be considered reliable as its calculated re-normalised unit weight error (RUWE) value is 1.148, below the recommended limit of 1.4. The \Gaia~eDR3 parallax of the B+C pair is $5.219\pm0.072$, consistent with that of A+D \citep{2021A&A...650C...3G}.
Other stellar parameters suggested by X-SHOOTER observations of \citet{2017A&A...604A.127M} are $T_\mathrm{eff}=5110$\,K, $L = 5.1\,L_\sun$, and $A_\mathrm{V} = 1.5$\,mag
. We rescaled the luminosity value with the \Gaia~eDR3 distance, and use $L = 7.2\,L_\sun$ in the following study. ALMA observations outlined a disk around DI Cha A, although the disk was unresolved in these measurements.



DI~Cha~A was observed with the MID-infrared Interferometric instrument (MIDI) on the Very Large Telescope Interferometer ( VLTI) multiple times between 2006 and 2009. The spatial resolution of MIDI permits a study of disk structure on $\sim$au scale. DI Cha A was included in the sample of 82 young stellar objects whose archival MIDI observations were analysed by \cite{2018A&A...617A..83V}. DI Cha A stood out of the sample as one of those few objects where an analytic temperature gradient model with an inner cavity produced the best fit to the data, in contrast to a continuous disk model. 

This result prompted us to initiate a more detailed disk study specifically for DI Cha A, using additional measurements and methods as well. Here, we combine photometric and interferometric data to study the structure of the inner region of the disk, which may be the site of early giant planet formation. We intend to prove the existence of the inner cavity using more sophisticated modeling tools, check for possible time variability, and examine the dust composition in the disk. A special characteristics of DI Cha A is its relatively high mass (2.08~M$_{\odot}$) which places it  among the most massive T Tauri stars, representing a borderline case toward the low-mass Herbig Ae stars. The higher stellar luminosity may have effects on the circumstellar disk, which we can explore in our study.


\section{Data and reduction}\label{sec:Photometry}
\subsection{MIDI observations}

We used the interferometric data taken by the MIDI instrument on the VLTI. MIDI was a Michelson-interferometer, which combined the beams from two telescopes of the VLT, using half-reflecting plate optical combiner \citep{2003Ap&SS.286...73L}. It produced spectrally resolved interferometric measurements ($R=30$) in the  7.5--13~$\mu$m wavelength range, using various baselines and position angles. 
The obtained data contain low-resolution spectra of the full integrated emission of the source (total spectra); spectra of the emission components that remained spatially unresolved on the different interferometric baselines (correlated spectra); the ratio of the correlated-to-total spectra, called visibility, that shows to what degree the source was spatially resolved; and differential phases, which are practically the phases of the visibility function, and indicates how centrally symmetric the source is \citep{2021arXiv211015556L}. In our study we used the measured visibility values in the N band at 10.7~$\mu$m to compare with the calculated visibility functions from modeling. 

\begin{figure}[!ht]
\includegraphics[width=\columnwidth]{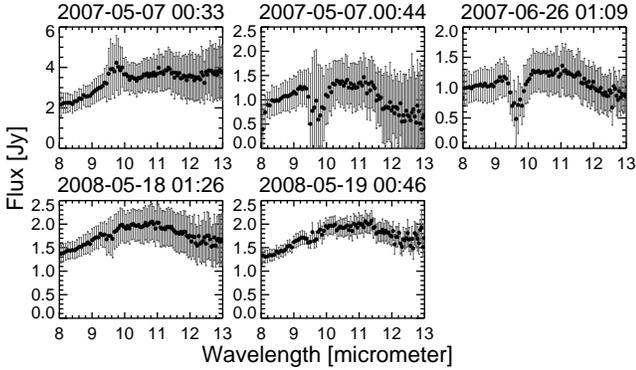}
\caption{MIDI total spectra of DI~Cha~A. For the list of observations used in the analysis we refer to Table~\ref{table:obs}.}
\label{fig:midi_total}
\end{figure}

\begin{figure}[!ht]
\begin{center}
\includegraphics[width=\columnwidth]{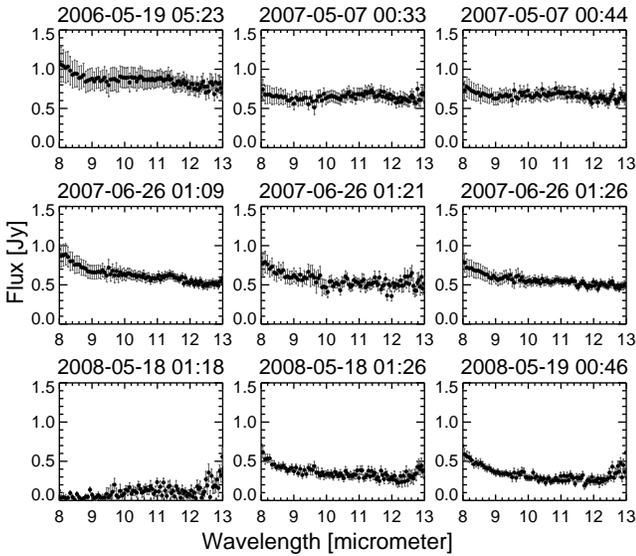}
\caption{MIDI correlated spectra of DI~Cha~A. For the list of observations used in the analysis we refer to Table~\ref{table:obs}.}
    \label{fig:midi_corr}
\end{center}
\end{figure}

\begin{figure}[!ht]
\begin{center}
\includegraphics[width=\columnwidth]{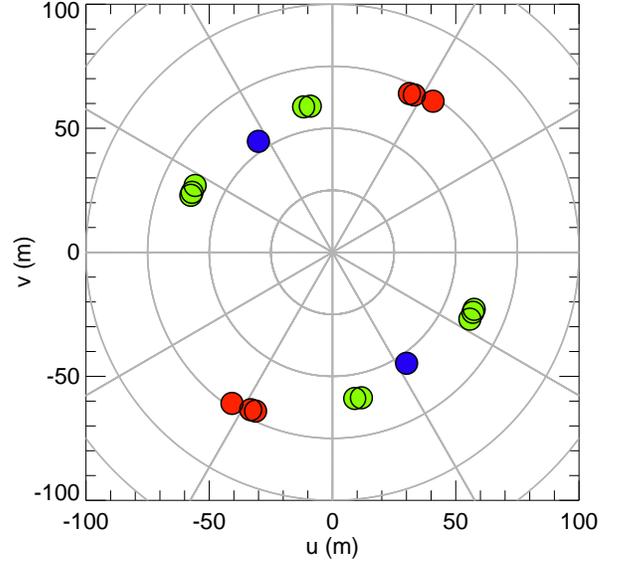}
\caption{The $uv$ space coverage of the MIDI observations. Blue color: data from 2006, green: 2007, red: 2008.}
    \label{fig:midi_uv}
\end{center}
\end{figure}

The MIDI observations of DI Cha A were carried out between 2006 May and 2008 May. The raw data set is available from the ESO public archive. We took the reduced and calibrated data from \cite{2018A&A...617A..83V}, which is a comprehensive compilation of MIDI observations of low- and intermediate-mass young stellar objects. For that paper, the data have been reduced with Expert Work Station (EWS) package 2.0 \citep{2004SPIE.5491..715J}, which  is  a  common tool for  MIDI  data  processing. The MIDI observations are listed in Table \ref{table:obs}. Some measurements were of lower quality (marked in Table \ref{table:obs}), and were not used in our analysis. At a few epochs only correlated spectrum measurements were performed, but if no technical issues were identified then these observations were also used. The resulting total and correlated spectra are shown in Figure~\ref{fig:midi_total} and Figure~\ref{fig:midi_corr}, respectively. The high dispersion of data points in several total flux spectra in Figure~\ref{fig:midi_total} at around 9.8~$\mu$m is due to atmospheric ozone contamination.
The $uv$ spatial coverage of the measurements is shown in Figure \ref{fig:midi_uv}.

\begin{table*}
\caption{Overview of VLTI/MIDI observations of DI Cha. $B$ is the projected baseline length, $\phi_B$ is the projected position angle of the baseline (measured from North through East), and $\tau_0$ is the atmospheric coherence time. The resolution is the approximate diameter of the beam at $10.7~\mu$m, converted to physical scale. Check mark denotes good quality measurement, x denotes that no measurement was made or we did not use it due to its lower quality.}
\label{table:obs}      
\begin{center}                          
\begin{tabular}{l l l l l l l l l l l}        
\toprule \toprule               
Date and time & Telescopes & $B$  &$\phi_B$ & Resolution  & Seeing & $\tau_0$ & Airmass & Program ID & \multicolumn{2}{c}{Quality} \\ 
(UTC) & & (m) & ($^\circ$) & (au) & (\arcsec) & & & & Total & Correlated \\
\midrule
2006-05-19 05:23 & U1U3 & 53.9 & 123.9 & 4.0 & 1.07 & 2.01 & 2.28 & 077.C-0662(A) & x & \checkmark\\
2007-05-07 00:33 & U3U4 & 59.6 & 98.7 & 3.7 & 0.72 & 1.82 & 1.66 & 079.C-0101(A) & x & \checkmark\\
2007-05-07 00:44 & U3U4 & 59.8 & 101.3 & 3.6 & 0.71 & 1.89 & 1.66 & 079.C-0101(A) & x & \checkmark\\
2007-06-26 01:09 & U3U4 & 61.9 &  154.2 & 3.5 & 1.03 & 1.35 & 1.92 & 079.C-0101(A) & x & \checkmark \\
2007-06-26 01:21 & U3U4 & 61.9 & 157.0 & 3.5 & 1,17 & 1.19 & 1.95 & 079.C-0101(A) & x & x\\
2007-06-26 01:26 & U3U4 & 61.9 & 158.2 & 3.5 & 1.19 & 1.17 & 1.96 & 079.C-0101(A) & x & \checkmark\\
2008-05-18 01:18 & U1U3 & 71.6 & 62.4 & 3.0 & 0.83 & 2.47 & 1.69 & 081.C-0020(A) & x & x\\
2008-05-18 01:26 & U1U3 & 71.2 & 64.0 & 3.1 & 0.83 & 2.46 & 1.70 & 081.C-0020(A) & \checkmark & x\\
2008-05-19 00:46 & U1U3 & 73.3 & 56.2 & 3.0 & 0.96 & 3.15 & 1.67 & 081.C-0020(A) & \checkmark & \checkmark\\
\bottomrule
\end{tabular}
\end{center}
\end{table*}

\subsection{PIONIER observations}
Interferometric data were also available from measurements obtained with the PIONIER instrument on the VLTI \citep{2011A&A...535A..67L}. PIONIER is a H-band (1.55--1.80~$\mu$m) instrument, that combines four beams either from the 8.2 m Unit Telescopes or from the 1.8 m Auxiliary Telescopes to provide the measurements of six visibilities and three independent closure phases with low spectral resolution ($R=40$). The PIONIER observations of DI~Cha~A were made on the nights 2012 March 4/5 (prop. ID: 088.C-0670(B), PI: F. Menard) and 2013 May 12/13 (proposal ID: 091.C-0570(A), PI: O. Absil). The reduced and calibrated data were taken from the Optical Interferometry DataBase, which is a public database of the JMMC\footnote{\url{https://www.jmmc.fr/english/tools/data-bases/oidb/}}.


\subsection{Spitzer monitoring observations at 4.5\,{\micron}}

DI\,Cha A was observed with the Infrared Array Camera \citep[IRAC,][]{fazio2004}  on the {\sl Spitzer} Space Telescope \citep{werner2004} serendipitously in the framework of a monitoring project (proposal ID: 60167, PI: P. \'Abrah\'am) targeting a nearby young star, WX\,Cha. A total of 14 observations were obtained between 2010 April 24 and May 7, in the 'warm mission' phase, with an average cadence of $\sim$1\,day. The IRAC camera was used in full-array mode with exposure times of 0.2\,s at 5 dithering positions. While WX\,Cha was observed in both available passbands at 3.6\,{\micron} and 4.5\,{\micron}, DI\,Cha A, which is located at $\sim$8\farcm5 from it, was covered only in the longer wavelength 
band. In addition, the latter source has measurements only at four dither positions typically, the exception is the 8th (at MJD of 55317.74250) and the last epochs, where there are observations at 5 and 3 positions, respectively.

In our analysis, we used the individual corrected basic calibrated data (CBCD) frames produced by the IRAC pipeline version S19.2.0. We performed aperture photometry using an aperture radius of 3\,pixels and a sky annulus between 12 and 20 pixels. To determine the centroid of the source we applied a first moment box centroider routine designed to work with IRAC data ('box\_centroider'\footnote{\url{https://irsa.ipac.caltech.edu/data/SPITZER/docs/dataanalysistools/tools/contributed/irac/box\_centroider/}}). To estimate the background, we used an iterative sigma-clipping method with a clipping threshold of 3$\sigma$. Then we performed array-location dependent and pixel phase photometric corrections by using the 'irac\_aphot\_corr' tool\footnote{\url{https://irsa.ipac.caltech.edu/data/SPITZER/docs/dataanalysistools/tools/contributed/irac/iracaphotcorr/}} provided by the Spitzer Science Center and an aperture correction using the appropriate correction factors from the IRAC Instrument Handbook v3.0.1\footnote{\url{https://irsa.ipac.caltech.edu/data/SPITZER/docs/irac/iracinstrumenthandbook/}}. The final photometric data and their uncertainties were computed as the mean and the error of the mean of the flux densities measured in the individual dither steps. The obtained photometry are listed in Table~\ref{tab:iracfluxes}.

	       

\begin{table}[!ht]
    \centering
    \begin{tabular}{c c c}
    \toprule \toprule 
        MJD & F$_{\mathrm{4.5{\mu}m}}$ & $\sigma$F$_{\mathrm{4.5{\mu}m}}$  \\
	     &    (Jy)   &  (Jy)   \\
	       
    \midrule
  55310.60828  &   1.866 & 0.031 \\
  55311.30709  &   1.892 & 0.017 \\
  55312.24676  &   1.882 & 0.032 \\
  55313.39812  &   1.883 & 0.041 \\
  55314.18235  &   1.865 & 0.015 \\
  55315.40390  &   1.825 & 0.024 \\
  55316.59242  &   1.827 & 0.027 \\
  55317.74250  &   1.842 & 0.026 \\
  55318.83573  &   1.855 & 0.041 \\
  55319.47407  &   1.852 & 0.016 \\
  55320.55676  &   1.795 & 0.042 \\
  55321.78600  &   1.793 & 0.023 \\
  55323.07712  &   1.803 & 0.044 \\
  55323.93855  &   1.782 & 0.045 \\
    \midrule
    \end{tabular}
    \caption{{\sl Spitzer}/IRAC photometry of DI\,Cha at 4.5\,{\micron}.} \label{tab:iracfluxes}
\end{table}

\subsection{TESS observations}

DI\,Cha has been observed by the Transiting Exoplanet Survey Satellite \citep[TESS,][]{ricker2015} both in 2019 as part of the Sector 11 and 12 campaigns and in 2021 during the Sector 38 and 39 observations. All measurements were performed in the full-frame image (FFI) mode with a cadence of 30 minutes in the first two sectors and 10 minutes in the the last two sectors. After we downloaded the calibrated FFI measurements from the MAST archive\footnote{\url{https://mast.stsci.edu}}, the photometry of DI\,Cha was extracted by performing a convolution-aided differential image analysis pipeline \citep[\texttt{qdlp-extract},][]{pal2020} that is largely based on the \texttt{FITSH} tool \citep{pal2012}. As part of the data reduction, we employed aperture photometry using a radius of 2.5\,pixels ($\sim$53{\arcsec}) and a background annulus between 5 and 10 pixels. The pipeline identifies and assigns flags to observations affected by significant stray light or momentum wheel desaturation. Because these points are of doubtful quality, we removed them from the data stream. The typical formal uncertainty of our 30 minutes and 10 minutes cadence photometry, is 0.17\,mmag and 0.30\,mmag, respectively. The reduced light curves are shown in Figure \ref{fig:TESS}.

\subsection{WISE observations}

In order to analyze the temporal behavior of DI Cha at mid-infrared wavelengths, we collected observations obtained by the Wide-field Infrared Survey Explorer \citep[WISE,][]{wright2010}. We used data obtained in the 3.4 $\mu$m (W1) and 4.6 $\mu$m (W2) bands  both in the cryogenic phase (2010 Feb and Aug, \citep{{10.26131/IRSA143}}), and in the NeoWISE mission (from 2014 until
the latest available data points in 2020). For the 2010 Aug epoch we used the WISE 3-band catalog \citep{10.26131/IRSA149}.
In order to derive W1/W2 magnitudes, we checked each single-exposure WISE measurement, and discarded those exhibiting high formal uncertainties, or when a PSF template fitting of the source's brightness profile was of poorer quality. We also realized that measurements offset from DI Cha's nominal coordinates by more than $\sim$2$''$ provided systematically fainter photometry; these data points were  discarded from the further analyses. The remaining data were averaged per season, producing continuous W1 and W2 light curves between 2014 February and 2020 July, including 14 epochs with approximately six months cadence. Since the source is bright at mid-infrared wavelengths, we applied saturation photometric bias corrections, following Sect.~II.1.c.iv.a. of the NeoWISE Explanatory Supplement\footnote{\url{https://wise2.ipac.caltech.edu/docs/release/neowise/expsup/}}. The typical uncertainty of the seasonal average data photometric points were $\sim$0.08 mag. The W1/W2 band NeoWISE photometric data are listed in Table~\ref{tab:wise}. The WISE magnitudes listed in Table~\ref{tab:Photometry} were computed from the 2010 Feb and 2010 Aug cryogenic measurement, by combining the results of the two epochs (in the W1 and W3 bands data at both epochs were available, while the W2 magnitude is from 2010 Aug, and the W4 magnitude is from 2010 Feb).

\begin{table}[!ht]
    \centering
    \begin{tabular}{ccccc}
    \toprule \toprule 
        MJD & W1 & $\sigma$W1 & W2 & $\sigma$W2 \\
	     &    (mag)   &  (mag) &    (mag)   &  (mag)  \\
    \midrule
56707.9  &  5.691  &  0.067  &  4.822  &  0.070  \\
56888.2  &  5.631  &  0.267  &  4.805  &  0.068  \\
57071.6  &  5.358  &  0.077  &  4.851  &  0.072  \\
57247.0  &  5.231  &  0.134  &  4.700  &  0.069  \\
57435.9  &  5.684  &  0.079  &  4.863  &  0.089  \\
57606.3  &  5.473  &  0.120  &  4.726  &  0.052  \\
57801.5  &  5.394  &  0.060  &  4.804  &  0.056  \\
57968.6  &  5.334  &  0.029  &  4.665  &  0.104  \\
58167.3  &  5.158  &  0.034  &  4.700  &  0.039  \\
58329.7  &  5.383  &  0.042  &  4.900  &  0.059  \\
58531.3  &  5.137  &  0.097  &  4.510  &  0.059  \\
58694.0  &  5.037  &  0.240  &  4.597  &  0.037  \\
58897.8  &  5.689  &  0.191  &  4.857  &  0.122  \\
59059.8  &  4.963  &  0.175  &  4.671  &  0.068  \\
    \midrule
    \end{tabular}
    \caption{NeoWISE photometry of DI\,Cha in the W1 (3.4~$\mu$m) and W2 (4.6~$\mu$m) bands. \label{tab:wise} }
\end{table}

\begin{table*}[h]
    \centering
    \caption{Photometric records for DI~Cha.}
    
    \label{tab:Photometry}
    \begin{tabular}{cccccc}
    \hline
    \hline
    Telescope/Catalog   &  Filter   &   Magnitude &   Flux~[Jy]  &   Wavelength~[$\mum$]     & Epoch \\
        \hline
    
Spitzer                &  MIPS:160       & $-3.580$ & $4.3$      & $155.9$ &  2005 \\
Spitzer                &  MIPS:70       & $-1.605$ & $3.4$      & $71.42$&       \\
Spitzer                &  MIPS:24       & $1.093$ & $2.61$      & $23.67$&        \\
Spitzer                &  IRAC:8.0 &      4.143  & $1.39$      & $7.87$&          \\
Spitzer                &  IRAC:4.5       & $4.997$ &  $1.8$      &  $4.49$&        \\
Spitzer                &  IRAC:3.6       & $5.356$ & $2.0$      & $3.54$&          \\
    \hline
    
WISE [*]               &  W1       & $5.387$ & $2.195$      &       $3.35$      &    2010/02 to 2010/08 \\
WISE [*]            &  W2       &$4.456$&   $2.153$    &     $4.6$       &   2010/08   \\
WISE [*]            &  W3       &$3.11$&   $1.663$     &     $11.56$       &   2010/02 to 2010/08   \\
WISE  [*]            &  W4       &$1.057$&  $3.129$   &       $22$      &   2010/02     \\
\hline
AKARI                &  L18W       & & $2.48$      & $18.39$ &           2007 \\
AKARI                &  S9W       & & $1.76$      & $8.61$ &          \\
    \hline
    2MASS                &  K$_s$       & $6.217$ & $2.20$      & $2.16$ &           2000 \\
    2MASS                &  H      & $6.941$ & $1.76$      & $1.65$ &         \\
    2MASS                &  J       & $7.818$ & $1.18$      & $1.24$ &           \\
\hline
    Gaia                &  Bp       &    $11.017$   & $0.3733$    &  $0.532$         &             2014/07 to 2016/05 \\
    Gaia                &  G        &  $10.194$     &    $0.3782$    &   $0.673$      &                \\
    Gaia                &  Rp       &        $9.316$&      0.3646    &      $0.77$  &             \\
    \hline
    SkyMapper           &     v     & $12.872$ &     $0.0347$         &     $0.411$      &   2014 --     \\
    SkyMapper           &     g     & $11.044$ &     $0.1426$         &      $0.524$      &       \\
    SkyMapper           &    r      & $10.268$ &     $0.3508$         &     $0.670$      &        \\

    SkyMapper           &     i     & $9.709$ &  $0.6223$         &         $0.790$      &   \\
    SkyMapper           &     z     & $9.460$ &  $0.7909$         &         $0.910$      &       \\
    \hline
    DENIS           &    DENIS:I  & $9.278$ &  $0.43$         &        $0.79$      &     1996 to 2001 \\
    DENIS           &    DENIS:J  &  	$7.872$ &    $1.225$         &      $1.24$      &      \\
    DENIS           &    DENIS:K  &  	$6.142$ &     $2.245$         &     $2.16$      &      \\
    \hline
    APASS           &    Sloan:g'      &  	$11.397$ &         $0.481$      &   $0.1003$         &  2011/02 to 2014/05    \\
    APASS           &    Sloan:r'      &  	$10.224$ &      $0.2954$         &    $0.623$      &        \\
    APASS           &    Sloan:i'      &  	$9.656$ &      $0.4985$         &    $0.764$      &        \\
    APASS           &    Sloan:z'      &  	$9.316$ &      $0.6693$         &     $0.906$      &       \\
     \hline
    GALEX                &  FUV       & 19.165 &   $0.0000783$      & $0.153$ &          2011   \\
    GALEX                &  NUV      & 17.219 &   $0.00047$      &  $0.231$ &          \\
    \hline
\multicolumn{6}{l}{[*] Average of the 2010 Feb (4-band cryogenic) and 2010 Aug (3-band cryogenic) WISE measurements.}\\
    \end{tabular}
\end{table*}

\section{Results}\label{sec:Results}
In this study we fit simultaneously the interferometric data and the SED. For the latter, we collected photometry from a wide range of instruments. As these data were observed at different epochs, first we need to check the time variability of the source at different wavelengths. 

\subsection{Short-term variability}

\begin{figure*}[!ht]
    \centering
    \includegraphics[width=\textwidth]{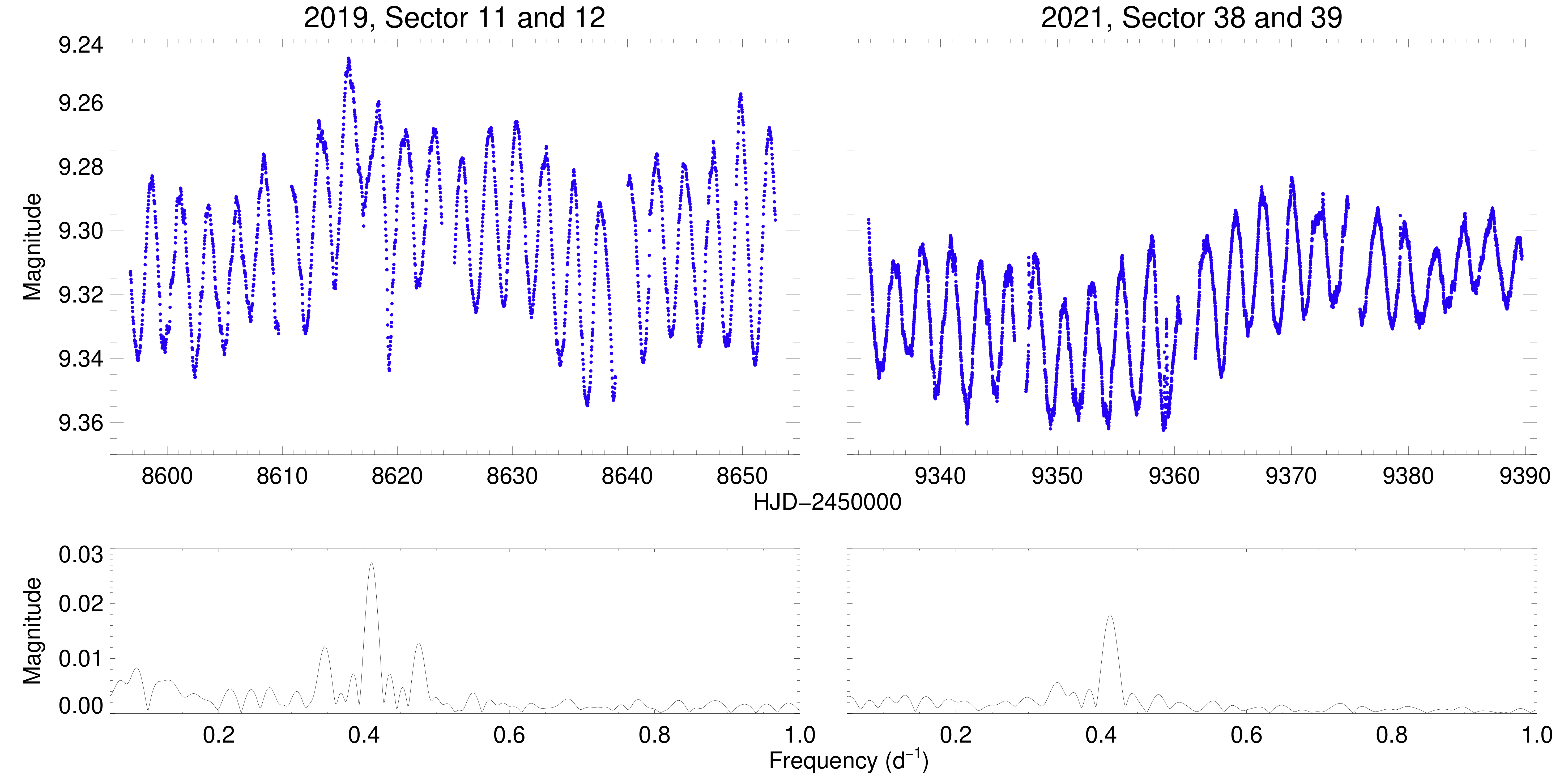}
    \caption{Top panels: The {\TESS} light curves of DI~Cha from 2019 (left) and 2021 (right). Bottom panels: The Fourier frequency spectra of the light curves.}
    \label{fig:TESS}
\end{figure*}

Information on daily, or even shorter timescales at optical wavelengths are provided by the TESS light curves, displayed in Figure~\ref{fig:TESS}. The source exhibited continuous low-level ($\Delta$mag$<$0.1 mag) variability in both 2019 and 2021. The most conspicuous feature is a regular periodic signal. In order to characterize it, we performed a frequency analysis of both light curves using the Period04 program \citep{2005CoAst.146...53L}, exploring periodicity in the 1–20 day range. The calculated Fourier frequency spectra are shown in the bottom panels of Figure \ref{fig:TESS}. The most significant peak in both 2019 and 2021 is at 0.41\,d$^{-1}$, corresponding to a period of 2.44 days. This period presumably represents the rotation of the star, as the rotation of a long-lived spot on the surface of the star may cause the observed magnitude fluctuations. While the light curves may suggest some longer periods, the TESS time series are limited in length to determine their significance.
 
We examined whether the observed variability could be due to contamination from neighboring sources. On the CCD camera of TESS, one detector pixel corresponds to 21$''$ also on the sky. An area of this size includes the M-spectral type companions of DI Cha A. However, the high-resolution measurements of Gaia eDR3 \citep{2016A&A...595A...1G,2018A&A...616A...1G} imply that the magnitude of the B+C secondary is only $G = 15.32$\,mag, thus, it contributes less than $~1\%$ to the brightness of DI Cha A ($G=10.21$\,mag), thus the contamination is negligible. 

\begin{figure}[!ht]
    \centering
    \includegraphics[width=0.48\textwidth]{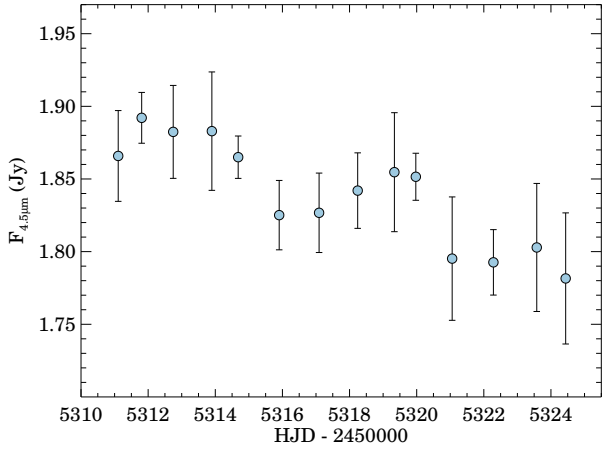}
    \caption{IRAC light curve of DI\,Cha.}
    \label{fig:spitzer_lc}
\end{figure}

In the infrared domain, Figure~\ref{fig:spitzer_lc} shows the 4.5\,$\mu$m Spitzer IRAC light curve of DI\,Cha. The average formal statistical uncertainty of the individual data points is 2\%, while the peak-to-peak flux change is $\sim$6\%. According to the IRAC Instrument Handbook, the repeatability of the measurements of individual sources is typically less than 1.5\%, suggesting that the observed fading trend is significant, thus the source is variable in the infrared on daily--weekly timescales. 
 

\subsection{Long-term variability}\label{subsec:long}

\begin{figure*}[!ht]
    \includegraphics[width=\textwidth]{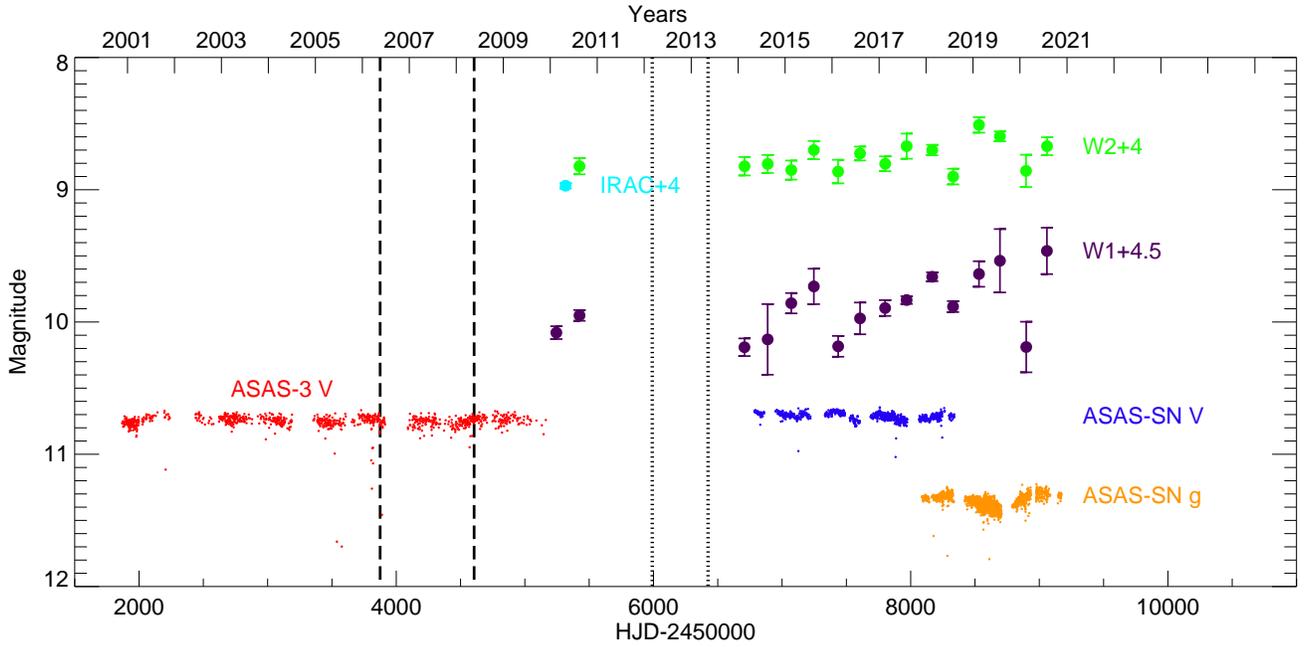} 
    \caption{The ASAS and WISE light curves of DI~Cha. The vertical dashed lines indicate the epochs of the first and last MIDI observations. The dotted line is same to the PIONIER observations. }
    \label{fig:ASAS}
\end{figure*}

A long series of optical  measurements of DI Cha is available from the All Sky Automated Survey (ASAS). One part of the light curve comes from the ASAS-3 system, which used two wide-field telescopes with 8.8x8.8 deg covering, equipped with V and I filters \citep{pojmanski2002,pojmanski2003}. The system also has one narrow-field telescope, which use I filter with 2.2x 2.2 deg covering. Each telescope equipped with a 2048x2048 pixels resolutions CCD camera. Another part of the measurements comes from the All-Sky Automated Survey for Supernovae (ASAS-SN). This program used 24 telescopes with 14 centimeters diameter, and its CCD camera could reach the 7.8 arcseconds resolution \citep{shappee2014,kochanek2017}. The observed light curves of the survey from DI Cha are shown in the Figure \ref{fig:ASAS}. Neither the $V$ band nor the $g$ band measurements show significant long-term annual time variability. The standard deviation of the ASAS-3 $V$ band magnitudes is 6.3\%, and in the case of the ASAS-SN $V$ band magnitudes is  2.7\%. The standard deviation is similar for the ASAS-SN $g$ band as well, its value is 4.6\%. 

We also retrieved optical photometric information of DI~Cha from Gaia eDR3 \citep{2021A&A...650C...3G}, which is combined information from the observations during the DR3 period (2014 July 25 to 2017 May 28), while the individual results are not yet public. The mean $G$ magnitude of the 405 good observations is 10.21\,mag, with flux accuracy of $(f/{\delta}f)_\mathrm{G}=954.3$. Following the method of \citet{2017MNRAS.467.2636D}, we estimated the flux scatter of multi-epoch Gaia observations as $\Delta{f}/f = \sqrt{N_\mathrm{obs}}\delta{f}/f =2.1\%$. Therefore, DI~Cha had only a weak variability in $G$ band during the ${\sim}34~$ months of Gaia DR3 observations.
Performing a similar study on the B+C components, the mean $G$ magnitude of the 349 good observations is 15.32\,mag. The flux accuracy is 472.7, and the estimated flux scatter is $3.9\%$.

\begin{figure}[!ht]
    \includegraphics[width=\columnwidth]{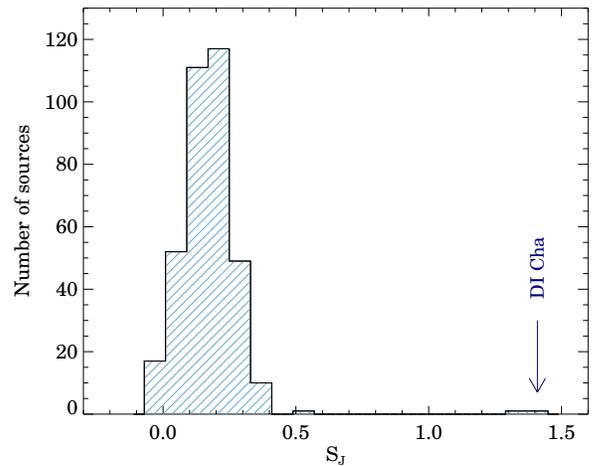} 
    \caption{Histogram of the Stetson J variability index, based on 357 reference stars. The $S_J$ value for DI~Cha is marked by a blue arrow.}\label{fig:stetson}
\end{figure}

%
In order to study the potential annual variability of DI~Cha at mid-infrared wavelengths, in Figure~\ref{fig:ASAS} we overplotted the NeoWISE photometric data. The light curves, in particular the W1 curve, seem to exhibit variability, most notably a gradual brightening trend over the whole covered period, with additional temporary fluctuations of the individual data points. To decide whether the magnitude changes are significant, we performed a statistical analysis, following the idea presented in \citet{moor2021}, using the 14 W1 and W2 NeoWISE data points. We computed the correlation-based Stetson J-index \citep[$S_J,$][]{stetson1996,sokolovsky2017}, and compared it with $S_J$ values derived for a sample of normal stars of similar brightness and located at similar ($\pm$5{\degr}) ecliptic latitudes to DI~Cha (${\beta}=-77\fdg6)$. The latter criterion was dictated by the fact that the mapping coverage (and thus the resulting sensitivity) of WISE depends on the ecliptic latitude. Stars in the reference sample were selected from the Hipparcos catalog in the A--K spectral type range. In total we collected 357 reference stars. Figure~\ref{fig:stetson} shows the histogram of the $S_J$ values. The majority of the data shows a Gaussian distribution, with a centroid of 0.17 and $\sigma=0.09$. The Stetson J-index of DI~Cha, $S_J=1.41$, is offset from the centroid at ${\sim}13.7\sigma$ level. The large $S_J$ value implies that the variations in the W1 and W2 light curves are highly correlated and the observed brightening trend in the DI~Cha data is significant. In the sample there is only one additional star with similarly large Stetson J-index ($S_J=1.32$), and we found that it is also a protoplanetary system, T~Cha, harboring a transitional disk \citep[e.g.,][]{hendler2018}.

Our results demonstrate that on both short and long timescales,  DI~Cha seems to be constant in the optical, apart from a periodic rotational signal, but exhibits statistically significant variability in the mid-infrared. For a discussion of these findings we refer to Section~\ref{sec:variability}.

\subsection{Spectral energy distribution}
\label{SED_anal}

\begin{figure}[!ht]
    \centering
    \includegraphics[width=\columnwidth]{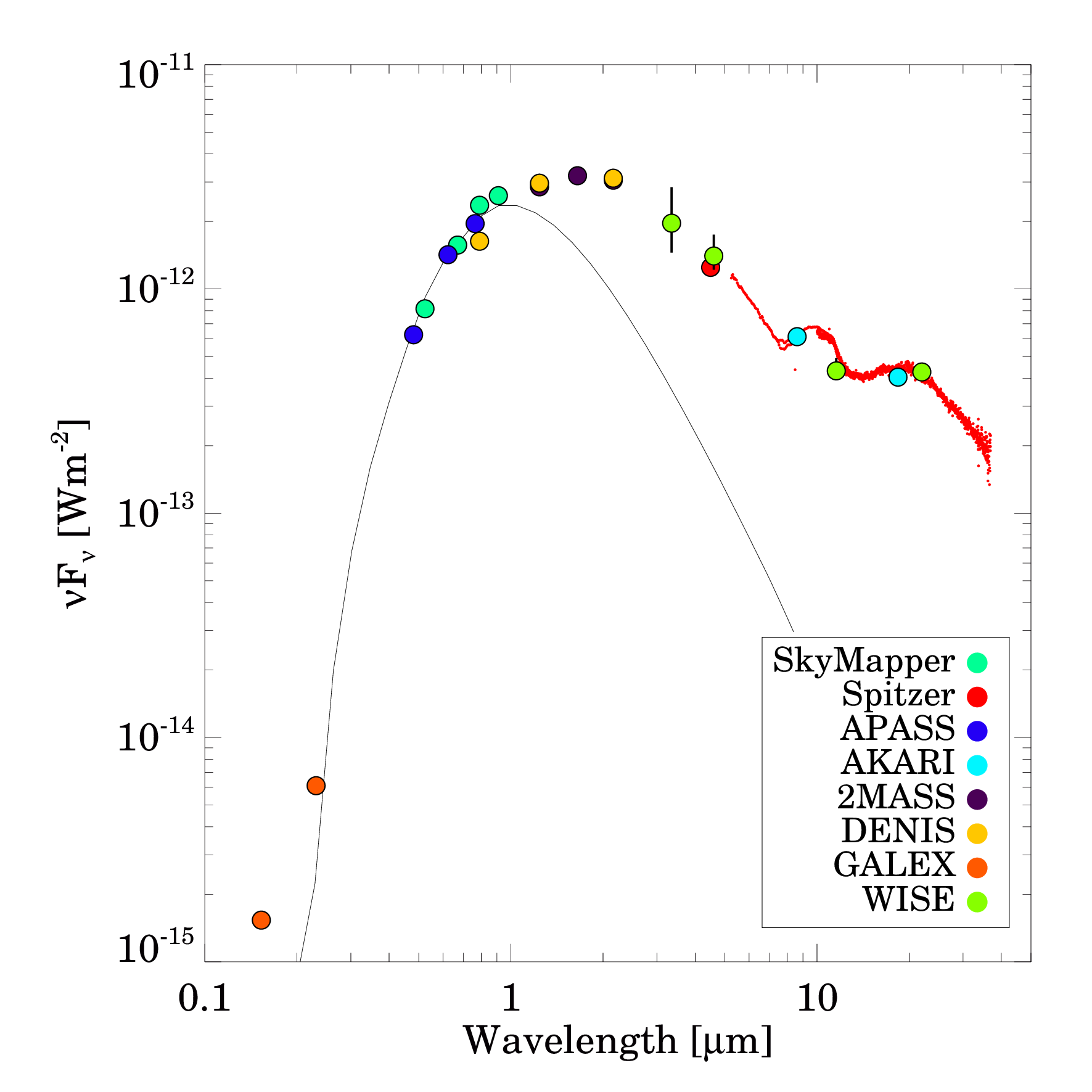}
    \caption{The observed SED of DI~Cha. The photometric data came from the following public archives: Spitzer; \cite{2008ApJ...675.1375L}, WISE; \cite{2012yCat.2311....0C}, AKARI; \cite{2010A&A...514A...1I} , 2MASS; \cite{2003yCat.2246....0C}, Gaia; \cite{2018A&A...616A...1G}, SkyMapper; \cite{2018PASA...35...10W}, DENIS; \cite{2005yCat.2263....0D}, APASS; \cite{2018AAS...23222306H},
    GALEX; \cite{2011Ap&SS.335..161B}. The vertical bar connecting the extreme values of the WISE light curves are plotted to illustrate the variability.} The black curve is the blackbody spectrum of the star distorted by reddening with $A_\mathrm{V} = 2.1$\,mag. (This value more than the interstellar extinction, as circumstellar extinction also has a contribution.)
    \label{fig:sed_full}
\end{figure}

Since no optical variability of DI Cha A was observed, we could combine optical photometric data obtained at various epochs to construct the short wavelength part of the SED. The data are summarized in Table~\ref{tab:Photometry}. At infrared wavelengths, the source was shown to vary on different timescales. Nevertheless, to construct an approximate SED, we combined 2MASS data (obtained in the year 2000) with Spitzer spectra (2006), and Spitzer photometric results (2005), to build the infrared part of the SED. The Spitzer spectra, including a low-resolution spectrum in the 5.2--14$\,\mu$m wavelength range and a high-resolution spectrum at 14--37$\,\mu$m, were downloaded from the CASSIS database \citep{2011ApJS..196....8L, 2015ApJS..218...21L}. The resulting SED is shown in the Figure \ref{fig:sed_full}. While the used infrared observations were not far in time from the MIDI measurements (2005--08), the absolute N-band flux level of the MIDI observations and that of the SED in Figure~\ref{fig:sed_full} might be different due to the variability. 

The SED exhibits a broad peak with a maximum between 1 and 2 $\mu$m, which represents the reddened photosphere of the star. The data points at longer wavelengths are above this photospheric level, revealing an infrared excess, that we attribute to the thermal emission of a circumstellar disk. The disk spectrum, in addition to a steady decline with wavelength as predicted by models of flared disks, shows two spectacular features. The first is a pair of two peaks at $\sim$10 and $\sim$18~$\mu$m, emitted by micrometer-sized silicate grains. The other is a shallow bump between $\sim$1 and $\sim$8~$\mu$m (better visualized in Figure~\ref{fig:sed_mir}), similar to the NIR bump often seen in Herbig Ae/Be stars peaking around 2-3~$\mu$m \citep{2001ApJ...560..957D,2001A&A...371..186N}. The usual explanation for such a bump is a vertical puffed-up rim situated at the inner edge of the protoplanetary disk \citep{2001ApJ...560..957D}. 


\section{Modeling the circumstellar structure}

Aiming to determine the structure and properties of the circumstellar disk, we fitted the measured interferometric and photometric data performing both geometric and radiative transfer modeling. As a first step, we adopted a geometric model, where we pre-defined a brightness distribution for the source. Based on the assumed brightness distribution, we performed the interferometric modeling in two ways: using either 1D Gaussian (Section~\ref{subsec:gaussmod}) or 1D temperature gradient models (TGM, Section~\ref{subsec:tgm}). After generating the image of the source with the given distribution, we calculated the visibility values of the model with a discrete Fourier transform at the $uv$-coordinates of the data following \cite{2015A&A...581A.107M} and \cite{2017A&A...604A..84V}. Then, we compared the model with the observations by estimating the $\chi^2$. To minimize the $\chi^2$ of the fitting we used a Python implementation of Goodman \& Weare’s Markov chain Monte Carlo (MCMC) ensemble sampler, called emcee \citep{2013PASP..125..306F, 2013ascl.soft03002F}. We ran chains of 10,000 steps, with 32 walkers. The first 1000 steps in the MCMC chain were discarded when calculating the best-fit values and errors. After fixing the inclination and the position angle of the disk by the geometric models, and obtaining some initial hints for the true spatial distribution of the circumstellar matter, we performed a 3D radiative transfer modeling in Section~\ref{subsec:radmc}.


\subsection{Gaussian model}
\label{subsec:gaussmod}

The MIDI data were modeled in a two-step procedure. First, we prepared a Gaussian model at 10.7~$\mu$m for the purpose of determining the position of the disk, and fitted the correlated fluxes.
\begin{figure*}[!ht]
    \centering
    \includegraphics[width=0.75\textwidth]{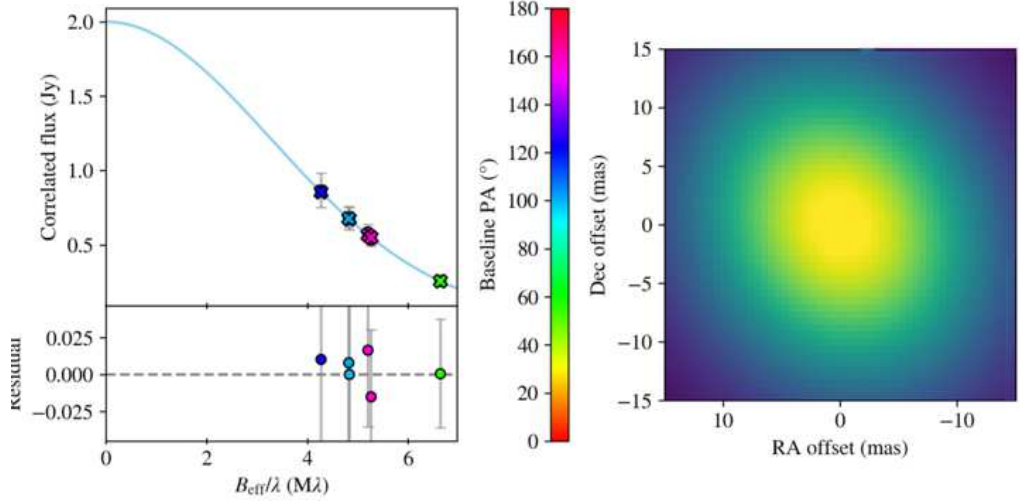}
    \caption{Results of the Gaussian model fit to the MIDI data at 10.7$\,\mu$m.}
    \label{fig:midi_gaussian}
\end{figure*}
The result of this Gaussian fitting is shown in Figure \ref{fig:midi_gaussian}. The resulting inclination is $29.54^{+4.36} _{-6.47}\,{\rm deg}$, and the position angle (PA) of the disk is $30.7^{+6.7} _{-5.8}\,{\rm deg}$. The full width at half maximum (FWHM) of the disk on 10.7 $\mu$m is 22.98 mas.

As the second step of the modeling, we determined especially the FWHM of the brightness  profile of the disk at different wavelengths. In these cases we used the $i$ and $PA$ values as fixed parameters, since our main goal was to detect the wavelength-dependent trend in the radiation areas. The resulting sizes for different wavelengths are shown in Table~\ref{tab:gaussian}. In the 10-11 micron range the obtained FWHM was $21.05^{+0.30} _{-0.29}$ mas, which is almost comparable to the results of \citet{2015A&A...581A.107M}. Their fitting result on 10.7 $\mu$m was a half-light radius of $14.1^{+1.0} _{-1.0}$  mas, assuming a face-on inclination. Taking into account the high uncertainties of the optical interferometric data the results can be considered consistent. 

\begin{table}[!ht]
    \centering
    \begin{tabular}{c c c}
    \toprule \toprule 
        Wavelength range &  \multicolumn{2}{c}{Fitted FWHM} \\
        & (mas) & (au) \\
    \midrule
        1.65 $\mu$m & $0.98^{+0.01} _{-0.01} $ & $0.18^{+0.01} _{-0.01} $ \\
        8--9 $\mu$m & $15.30^{+0.33} _{-0.32} $ & $2.92^{+0.06} _{-0.06} $ \\
        9--10 $\mu$m & $18.61^{+0.33} _{-0.32} $ & $3.55^{+0.06} _{-0.06} $ \\
        10--11 $\mu$m & $21.05^{+0.30} _{-0.29} $ & $4.02^{+0.06} _{-0.06} $ \\
        11--12 $\mu$m & $23.27^{+0.31} _{-0.31} $ & $4.44^{+0.06} _{-0.06} $ \\
        12--13 $\mu$m & $25.86^{+0.36} _{-0.35} $ & $4.94^{+0.07} _{-0.07} $ \\
    \midrule
    \end{tabular}
    \caption{Results of the Gaussian model fitting for the each wavelength range.}
    \label{tab:gaussian}
\end{table}

\begin{figure*}[!ht]
\centering
    \includegraphics[width=0.75\textwidth]{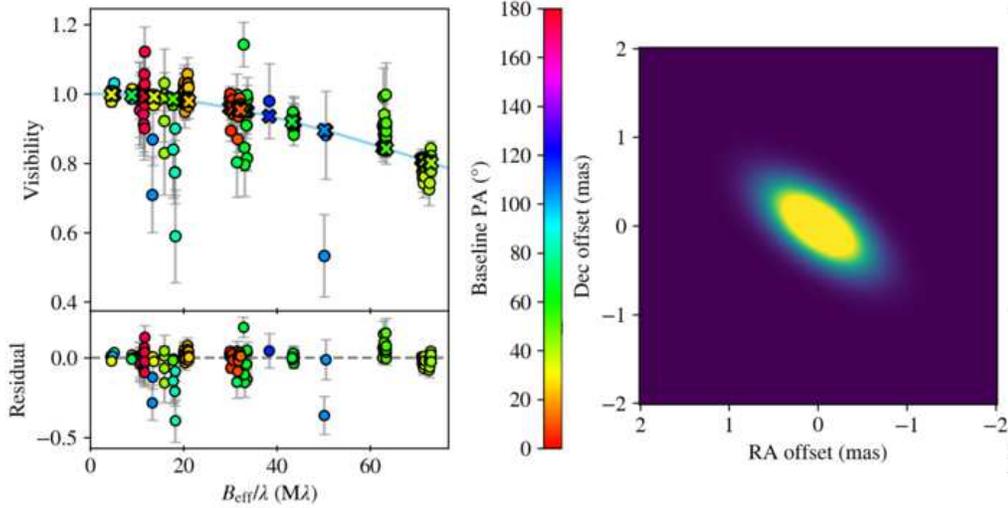}
    \caption{Results of the Gaussian model fitting to the PIONIER data at 1.65$\,\mu$m.}
    \label{fig:pionier_gaussian}
\end{figure*}

We also constructed a Gaussian brightness distribution model to reproduce the PIONIER data  at 1.65~$\mu$m. In this fit, both the position parameters of the disk ($i$, PA) and the FWHM were fitted parameters. The radiation contribution of the star was taken into account in the modeling. The results are shown in Figure \ref{fig:pionier_gaussian}. The obtained inclination is $61.96^{+7.42} _{-11.77}\,{\rm deg}$, and the PA is $53.3^{+4.6} _{-3.0}\,{\rm deg}$. The resulting FWHM is $1.01^{+0.01} _{-0.01}$ mas.

The inclination and PA values derived from the PIONIER observations differ from the corresponding numbers based on the MIDI data. However, at 1.65~$\mu$m the source was only marginally resolved by PIONIER (all visibility values are above 0.8), thus the derived parameters on the disk geometry are more uncertain. In the subsequent analyses we will adopt the inclination and PA values based on the MIDI observations. The 1.65~$\mu$m FWHM obtained using the i and PA values of MIDI is shown in Table~\ref{tab:gaussian}.
 
\begin{figure}[!ht]
    \centering
    \includegraphics[width=0.45\textwidth]{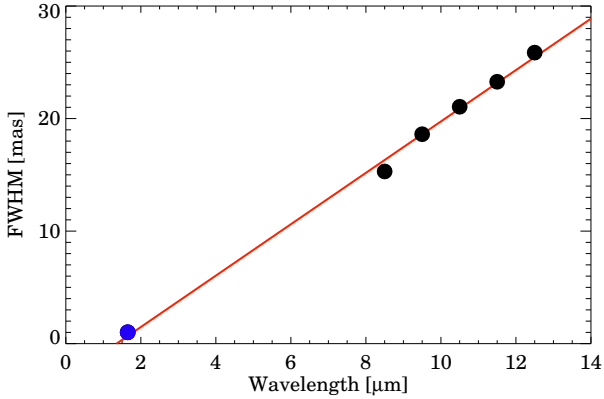}
    \caption{Results of the Gaussian model fitting for the each wavelength ranges. The black points marks the MIDI fitting results, the blue point marks the PIONIER fitting result.}
    \label{fig:temp_profile}
\end{figure}

The wavelength dependence of the size of the emitting area carries information on the temperature profile of the disk. The FWHM values from Table~\ref{tab:gaussian} are plotted in Figure \ref{fig:temp_profile}, together with a straight fit fitted to the area sizes. The good linear fit shows that the dimensions increase monotonically in proportion to the wavelength. This is a clear sign that the temperature of the material gradually decreases moving farther from the star. Thus, by examining a disk area with a larger angular diameter, we expect to observe more longer wavelength radiation. 

\subsection{Temperature gradient model}
\label{subsec:tgm}

We also fitted a temperature gradient model (TGM) to the interferometric data. We kept the inclination and the position angle fixed to the values determined from the Gaussian fits.
In the TGM model, as described in \cite{2017A&A...604A..84V}, 
the temperature ($T$) is a function of radius ($r$) as
\begin{equation}
T\left(r\right) = T_\mathrm{sub}\left(\frac{r}{R_\mathrm{sub}}\right)^{-q},
\end{equation}
where $T_\mathrm{sub}$ is the dust sublimation temperature, fixed at $1500$~K. $R_\mathrm{sub}$ is the sublimation radius, calculated from the known $L_\star$ luminosity of the central star:
\begin{equation}
R_\mathrm{sub} = \left(\frac{L_\star}{4\pi \sigma T_\mathrm{sub}^4}\right)^{1/2}.
\end{equation}
With $L_\star = 7.2$~L$_\sun$ \citep[][rescaled with the \Gaia~eDR3 distance]{2017A&A...604A.127M}, the sublimation radius is $R_\mathrm{sub} = 0.185~\mathrm{au}$.

\begin{figure*}[!ht]
\centering
    \includegraphics[width=0.75\textwidth]{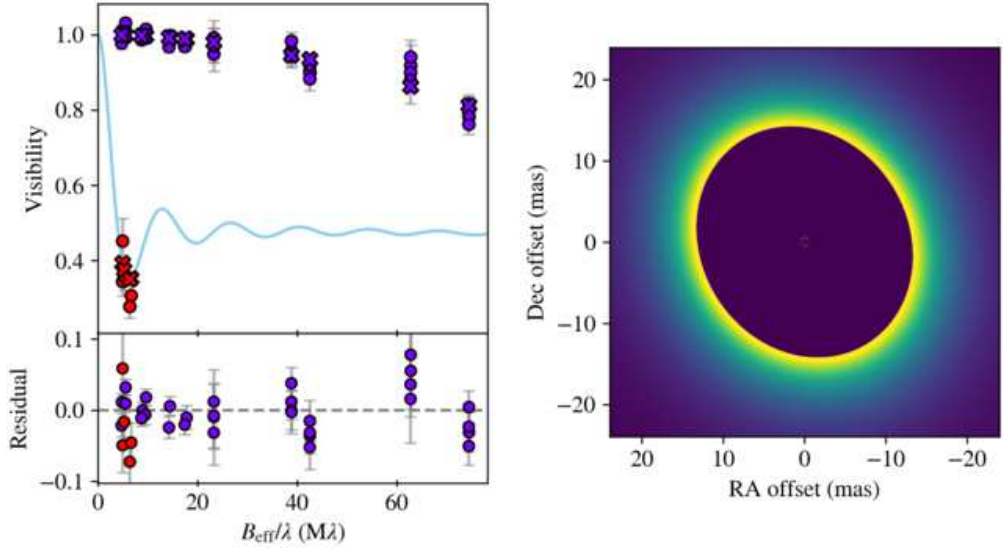}
    \caption{Results of the temperature gradient model fitting to the PIONIER and the MIDI data at 10.7$\,\mu$m.}
    \label{fig:pionier_midi_temp_grad}
\end{figure*}

In our modeling, the MIDI and PIONIER measurements were fitted simultaneously. In order to match both data sets, a two-component geometry was assumed, as was hinted by the SED analysis in Sect~\ref{SED_anal}. In this setup the disk has an inner cavity, as was already proposed by \cite{2017A&A...604A..84V}, but we additionally included a narrow ring in the vicinity of the star. The inner ring and the outer disk are separated by a gap. The model has four free parameters: $q$, which is the power-law exponent of the temperature profile, the $R_\mathrm{in}$ inner radius of the disk, the $R_\mathrm{gap,in}$ inner radius of the gap, and the $R_\mathrm{gap,out}$ outer radius of the gap. The results are shown in Figure \ref{fig:pionier_midi_temp_grad}, and the model parameters are listed in Table~\ref{tab:temp_grad}.
Our inner radius (0.099 au) is significantly smaller than the $1.6^{+0.4} _{-0.2}$ au value of \cite{2018A&A...617A..83V}. The difference was probably caused by use of different geometry involving the PIONIER data as well as adoption
lower star brightness (7.2 $L_\sun$ instead of 16.6 $L_\sun$) in the present work. 

\begin{table*}[!ht]
    \centering
    \begin{tabular}{c c c c c c c}
    \toprule \toprule 
        \multicolumn{2}{c}{$R_\mathrm{in}$} & \multicolumn{2}{c}{$R_\mathrm{gap,in}$} & \multicolumn{2}{c}{$R_\mathrm{gap,out}$} & $q$ \\
    \midrule
    (mas) & (au) & (mas) & (au) & (mas) & (au) & \\
    $0.53^{+0.08} _{-0.00} $ & $0.099^{+0.02} _{-0.00} $ & $0.56^{+0.01} _{-0.00} $ & $0.106^{+0.002} _{-0.00} $ & $14.72^{+0.77} _{-0.70} $ & $2.78^{+0.15} _{-0.13} $ & $0.63^{+0.01} _{-0.02} $ \\
    \midrule
    \end{tabular}
    \caption{Results of the temperature gradient model fitting.}
    \label{tab:temp_grad}
\end{table*}

\subsection{Radiative transfer model}\label{subsec:radmc}

\begin{figure}[!ht]
    \centering
    \includegraphics[width=0.45\textwidth]{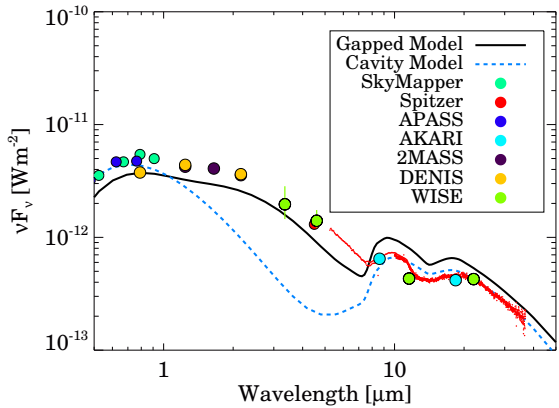}
    \caption{The spectrum of DI~Cha. Black curve: SED of the two component model (with gap). Blue dashed curve: SED of the one component (cavity) model.}
    \label{fig:sed_mir}
\end{figure}

We performed a radiative transfer modeling of DI Cha A using the RADMC-3D software package \citep{2012ascl.soft02015D}.  RADMC-3D calculates images and a SED for a given geometrical distribution of gas and/or dust, when viewed from a certain inclination angle. During the procedure, the program package also takes into account the contribution from scattering.
We computed models with two different geometrical setups. The first structure was a simple, one-component disk model (hereafter cavity model), suggested by the outcome of the temperature gradient modeling. At the start of our modeling, the inner radius of this disk was set to 0.91 au based on the TGM, and extended continuously to an outer boundary of 150 au. Generating the spectrum of this density distribution by radiative transfer modeling, however, produced an unsatisfactory fit to the measured data. For the best SED fit, we had to increase the inner boundary of the disk to 2.0 au. In this case, an appropriate model fitting could be achieved, except for the 1--8~$\mu$m wavelength range (Figure \ref{fig:sed_mir}, dotted line).

In order to reproduce also the near-infrared bump on the SED, we inserted an additional near-star disk component into the density distribution, creating a two-component disk model (hereafter gapped model). This geometry somewhat resembles that of the pre-transitional disks \citep{2010ApJ...717..441E}, however, in our setup the scale height of the inner disk could be high, almost forming a halo-like structure around the star. The presence of a halo-like inner component can also be found in several other modeling studies in order to successfully reproduce the measured near-infrared excess radiation \citep[e.g.,][]{2013A&A...555A..64M,2016A&A...587A..62K}.
In our best fitting model, the small inner component was located between 0.19 and 0.21 au, with a scale height of 0.5.  Since the inner disk made a large contribution to the mid-infrared radiation, in this second setup we had to extend the inner boundary of the outer disk to 3.0 au for the best fit. In order to the outer disk get enough radiation from the star, we built a puffed-up inner rim into our model. For this we used a simple geometry, there is a spherical shape inflated to a diameter of 0.3 au on the inner edge of the disc. In order to increase the radiation power around 10 microns, we also used a puffed-up inner rim for the cavity model, but that was a smaller sphere with a diameter of 0.1 au. 

\begin{figure*}[!ht]
\begin{center}
\includegraphics[width=0.31\textwidth]{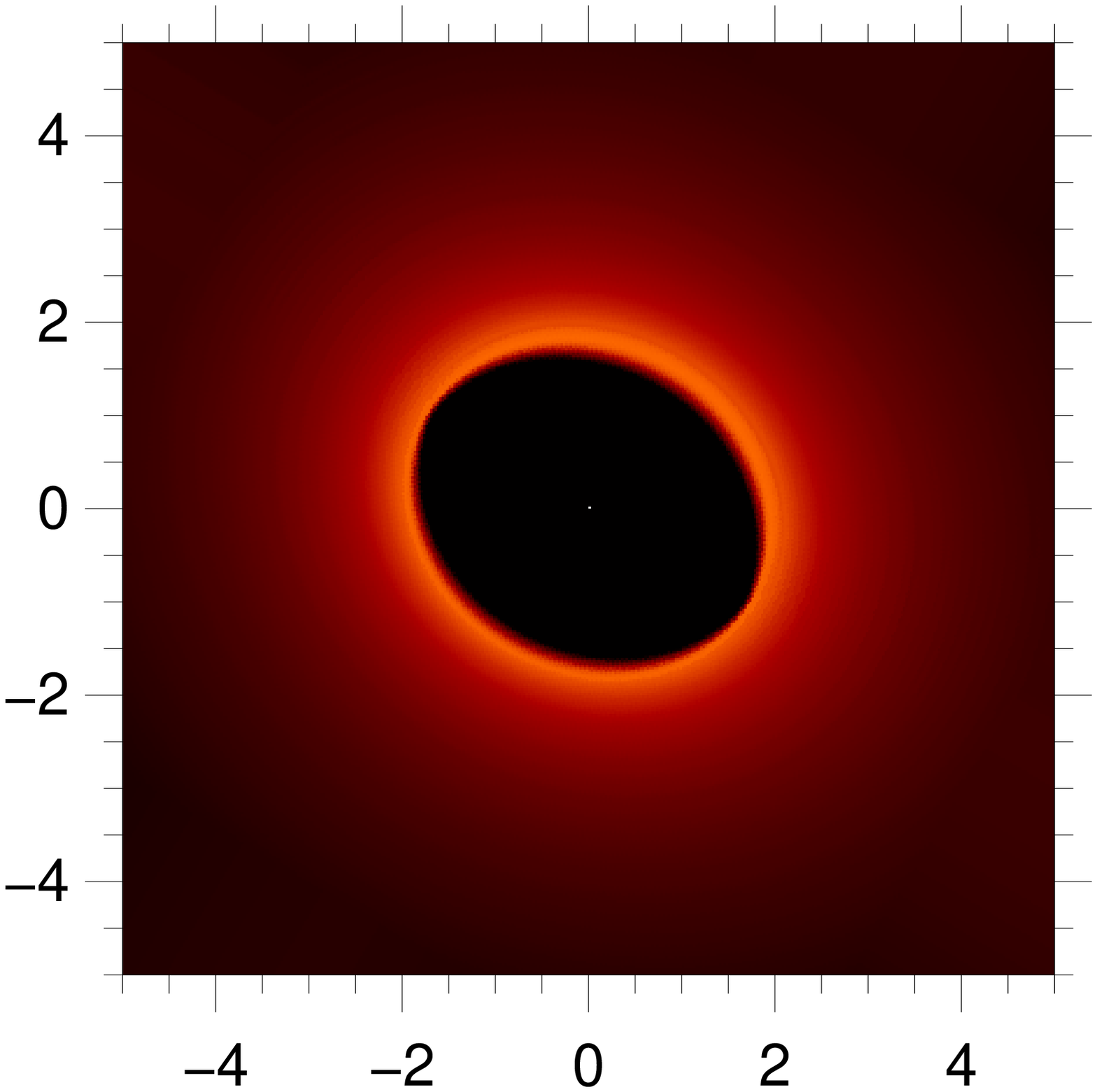}
\includegraphics[width=0.31\textwidth]{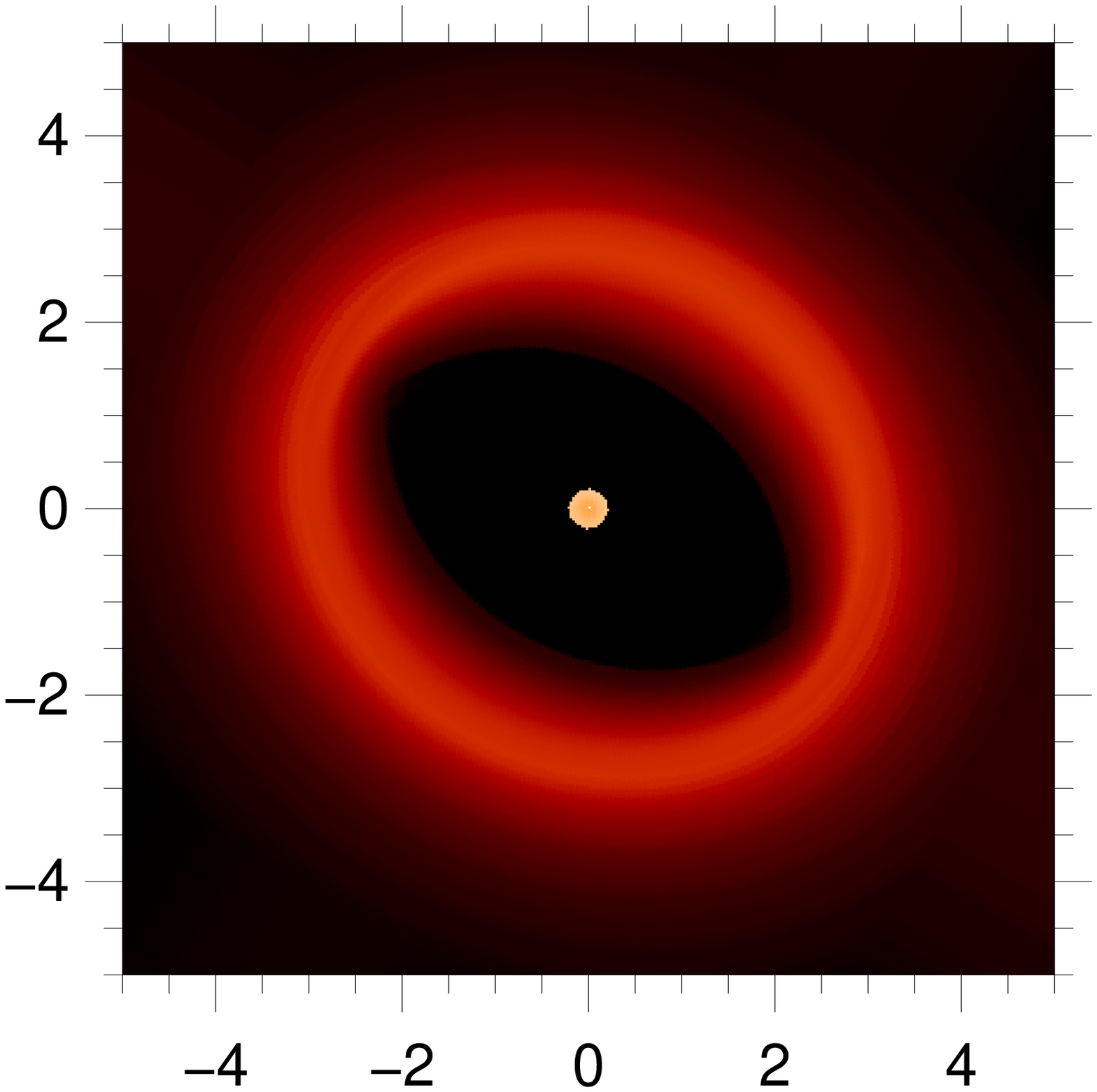}
\includegraphics[width=0.33\textwidth]{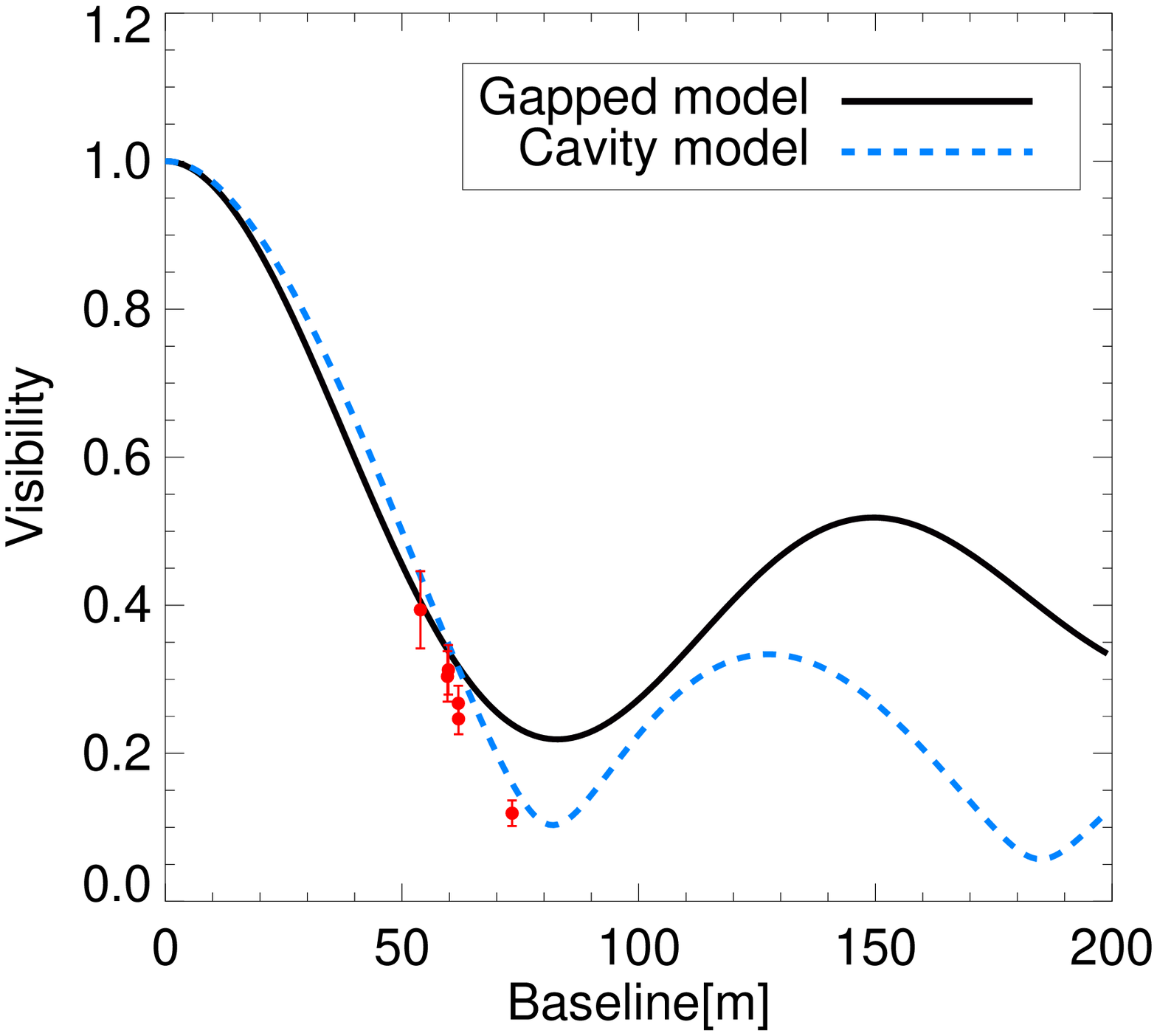}
\caption{Results from our radiative transfer models in the $N$ band (10.7~$\mu$m). Left: The image for a disk model with inner radius of $2.0$~au. Middle: A gapped model consisting of a tenuous inner dust between 0.19 and 0.21 au, and an outer disk with inner radius of $3.0$~au. Right: The visibility functions of both models, compared with the MIDI data (red dots). For the gapped model fitting $\chi^2$=0.1572 and for the cavity model fitting $\chi^2$=0.0558.}
    \label{fig:midivisibility}
\end{center}
\end{figure*}

\begin{figure}[!ht]
\begin{center}
\includegraphics[width=0.4\textwidth]{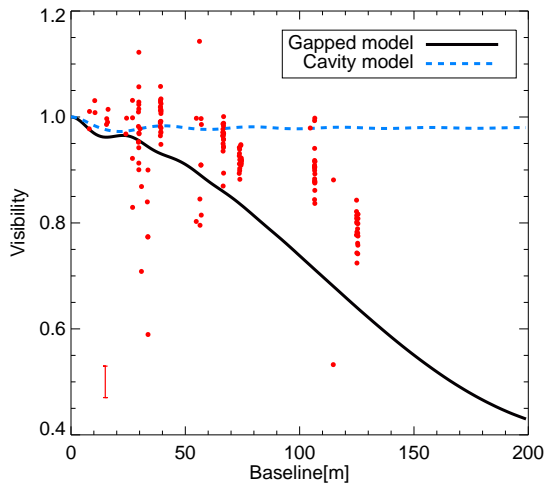}
\caption{The visibility functions of both models in the $H$ band (1.65~$\mu$m), compared with the PIONIER data (red dots). The error bar in the left corner reflects the average dispersion of the data. For the gapped model fitting $\chi^2$=2.66 and for the cavity model fitting $\chi^2$=2.44.}
    \label{fig:pioniervisibility}
\end{center}
\end{figure}

The parameters of the best fitting cavity and gapped models are listed in Table~\ref{tab:radmc}. Dust particle sizes between 0.01 and 1000 micrometers were used in both the inner and outer disk components. The number density of the larger particles decreased following a power-law in all cases, however with a larger exponent in the case of the outer disk. For the best fit, in the inner disk, we had to use significantly more processed particles, prescribed by setting a shallower grain size distribution. Based on our RADMC-3D modeling, we computed images of both models at 1.65$\,\mu$m and at 10.7$\,\mu$m. From the images we calculated the visibility functions of the models using Fourier-transformation. The resulting visibility functions were then compared with the visibility values from the VLTI/MIDI (Figure~\ref{fig:midivisibility}) and VLTI/PIONIER (Figure~\ref{fig:pioniervisibility}) data sets.

\begin{table}[ht!]
\caption{Parameters used in the best-fit models. Those in italics
were adopted from the literature and kept fixed during the modeling.}
\label{tab:radmc}
\begin{center}
    \begin{tabular}{l c}
    \toprule \toprule
    Parameters & Fitted value \\
    \midrule
    \textbf{System parameters} & \\
    \textit{Distance (d)} & 188.6 pc \\
    \textit{Inclination (i)} & 29.54$\degr$ \\
    \textbf{Stellar parameters} & \\
    \textit{Temperature} ($T_\mathrm{star}$) & 5110 K \\
    \textit{Radius ($R_\mathrm{star}$)} & 3.4 $\mathrm{R}_\mathrm{\odot}$\\
    \textit{Luminosity ($L_\mathrm{star}$)} & 7.2 $\mathrm{L}_\mathrm{\odot}$\\
    \textbf{Dust species} & Mass fraction \\
    Carbon & 20$\%$ \\
    Astrosilicate & 40$\%$ \\
    Crystal silicate (enstatite) & 40$\%$ \\
    Grain size range (for all species) & 0.01 - 1000 $\mu$m \\
    \textbf{Disk parameters of cavity model} & \\
    Inner radius ($R_\mathrm{indisk}$) & 2 au \\
    Outer radius ($R_\mathrm{disk}$) & 200 au \\
    Scale height ($\frac{h_\mathrm{disk}}{r_\mathrm{disk}}$) & 0.04 \\
    Flaring index ($\alpha_\mathrm{fl}$) & 0.1 \\
    Surface density at $R_\mathrm{indisk}$ ($\Sigma$) & $4\cdot 10^{-3}$ ${\mathrm{g}\cdot\mathrm{cm}^{-2}}$ \\
    Exponent of radial density profile (p) & -1.0 \\
    Power-law index of carbon grain size distribution & -3.0 \\
    Power-law index of astrosilicate grain size distribution & -5.0 \\
    Power-law index of enstatite grain size distribution & -3.5 \\
    \textbf{Inner disk parameters of gapped model} & \\
    Inner radius ($R_\mathrm{indisk}$) & 0.19 au \\
    Scale height ($\frac{h_\mathrm{disk}}{r_\mathrm{disk}}$) & 0.5 \\
    Flaring index ($\alpha_\mathrm{fl}$) & 0.0 \\
    Surface density at $R_\mathrm{indisk}$ ($\Sigma$) & $1\cdot 10^{-2}$ ${\mathrm{g}\cdot\mathrm{cm}^{-2}}$ \\
    Exponent of radial density profile (p) & -1.0 \\
    Power-law index of carbon grain size distribution & -3.0 \\
    Power-law index of astrosilicate grain size distribution & -2.0 \\
    Power-law index of enstatite grain size distribution & -2.5 \\
    \textbf{Outer disk parameters of gapped model} & \\
    Inner radius ($R_\mathrm{indisk}$) & 3 au \\
    Outer radius ($R_\mathrm{disk}$) & 200 au \\
    Scale height ($\frac{h_\mathrm{disk}}{r_\mathrm{disk}}$) & 0.02 \\
    Flaring index ($\alpha_\mathrm{fl}$) & 0.3 \\
    Surface density at $R_\mathrm{indisk}$ ($\Sigma$) & $3\cdot 10^{-3}$ ${\mathrm{g}\cdot\mathrm{cm}^{-2}}$ \\
    Exponent of radial density profile (p) & -1.0 \\
    Power-law index of carbon grain size distribution & -3.0 \\
    Power-law index of astrosilicate grain size distribution & -5.0 \\
    Power-law index of enstatite grain size distribution & -3.5 \\
    \midrule
    \end{tabular}
\end{center}
\end{table}

Figure~\ref{fig:sed_mir} shows the mid-infrared part of the SED, with the two model curves overplotted. While the ${\lambda}>8\,\mu$m range is slightly better reproduced by the simple disk model, the fit to the 1--8~$\mu$m spectral shape strongly prefers the gapped geometry, practically excluding the simple cavity model. While the match to the absolute flux level is not perfect, we recall that DI Cha A is variable at infrared wavelengths, which might explain part of the discrepancies. 

Figure~\ref{fig:midivisibility} shows the fitting results to the MIDI-based visibilities. While the cavity model is slightly more successful in reproducing the visibility values at all baselines, the gapped model is also consistent with most of the observations. Thus, we conclude that although the MIDI data prefer the cavity model from the two setups, they do not exclude the gapped model, too.

Although the PIONIER observations were not directly involved in the fitting procedure, we used them for a final consistency check. Their inspection revealed that the existence of an inner ring may be indicated by the PIONIER visibility data in Figure~\ref{fig:pioniervisibility}. In the cavity model the outer disk would not emit in the H-band. Thus the only emitter is the star itself, which is unresolved for PIONIER, predicting visibilities very close to 1.0. The observed visibilities, however, clearly drop at larger baselines, implying hot material outside the star. In our best fit, the scale height of the inner disk is rather large, $\sim$0.5. It resembles more a vertical cylindrical structure, or a spherical halo with a large polar opening, than a flat disk. The gapped model still does not fully reproduce the absolute level of the PIONIER visibilities, because the higher measured visibility values would indicate a more compact H-band source. In our model, however, the radius of the inner ring is very close to the silicate sublimation radius.
Invoking a population of higher temperature refractory grains located within the silicate sublimation radius might offer a solution.

Since the visibilies of the two models do not differ significantly on the short baselines, it is not possible to decide which model it supports based on the PIONIER data. Although the fitting suggests the presence of the inner, hot circumstellar matter, the $\chi^2$ values of the cavity model proved to be slightly more favorable.
In summary, we conclude that although the gapped model is not proven by the visibility data of either the MIDI or the PIONIER, they does not really rule out its correctness. Considering the result of the SED fitting, our gapped geometry is still strongly suggested. However, a perfect model reproducing all of the measurements can not be found which is probably related to the brightness variability also detected in the WISE data series (Section \ref{subsec:long}.).

\def\HeightImage{4.5cm}

\def\HeightImage{4.5cm}

\section{Discussion}\label{sec:Discussion}
\subsection{Geometry of the circumstellar environment}

In the previous section we demonstrated that while the MIDI visibilities slightly prefer the cavity model, the SED fitting support very strongly the gapped model, thus we advocate it as the geometry of the DI Cha A disk. 

In order to judge how typical or atypical the derived disk structure is among pre-main sequence stars, we refer to a study published in \citet{2018A&A...617A..83V}, who examined 82 disks around low- and intermediate-mass young stars based on VLTI/MIDI measurements. In that work, using the same temperature gradient models as we used here, they could fit most young stellar objects with a continuous model starting at the sublimation radius. There were only a few exceptions, including DI Cha, where the best fit was achieved by a model featuring an inner cavity. The authors examined the relationship between the luminosity of the central star and the estimated half-light radius of the disk in this sample. We identified DI Cha in these plots, and found that the high half-light radius of DI Cha A, 2.6 au, is rather high compared to the majority of other T Tauri stars. This result remains valid using L=$ 7{.}2 \ L_\odot$ what we adopted for the present work. Based on this finding, \cite{2018A&A...617A..83V} already suggested that DI Cha has an inner cavity in its disk, which was confirmed by our radiative transfer modeling, additionally arguing for the presence of a small inner disk. Since only 6 out of the 82 disks exhibite radii larger than the sublimation radius in the $N$ band, we can conclude that cavities and gaps on au-scale are relatively rare, and DI Cha A is a representative of this subclass. The inner gap was also detectable from our temperature gradient modeling, however, due to the fact that we also used $H$ band measurements during the fitting, the presence of a near-star, hot disk component also became apparent. Our results also suggest that this inner component may be located within the sublimation radis, suggesting the contribution of extreme heat-tolerant refractory particles in the disk. 

\subsection{Mineralogy}
We can derive information about the mineralogy of the dust component of the disk by analyzing the shape and strength of the 10$\,\mu$m silicate feature. Visually inspecting the correlated flux spectra in Figure~\ref{fig:midi_corr}, we concluded that while the measurements obtained with baselines $\leq$60~m exhibit weak silicate emission, such a peak is invisible in other observations taken on longer baselines. This result hints at a radial dependence of the silicate dust properties. It suggests that in the inner disk, which dominates the emission on longer baselines, the grain size distribution is skewed toward larger particles which do not emit the 10$\,\mu$m feature, hinting at an advanced grain growth process. The presence of an emission feature at shorter baselines, as well as in the Spitzer IRS spectrum (Figure~\ref{fig:sed_mir}), suggest, however, that the dust in the outer disk is different, and contains more small-size silicate particles.

\begin{figure}[h]
    \centering
    \includegraphics[width=\columnwidth]{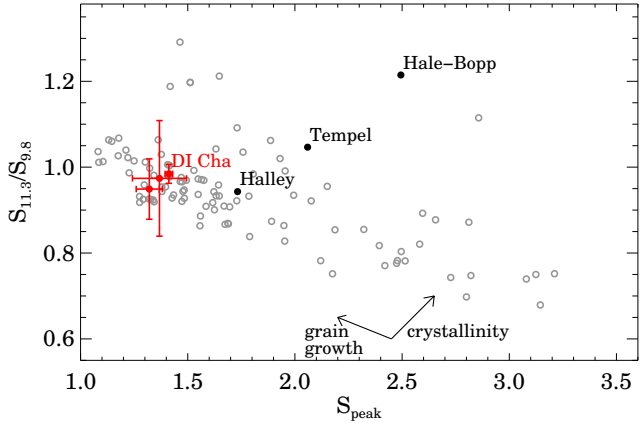}
    \caption{Ratio of the continuum subtracted and normalized 11.3/9.8~$\mu$m flux ratio versus the feature strength. DI Cha is marked with red dots. Gray circles mark T Tauri and Herbig Ae/Be stars. Black dots show solar system comets.}
    \label{fig:flux_ratio}
\end{figure}

To answer whether grain growth has also happened in the outer disk, we examined the spectral shape of the silicate emission feature following an analysis similar to that in \cite{2020ApJ...895L..48K}, who studied the relationship between the continuum subtracted and normalized 11.3/9.8$\,\mu$m flux ratio and the silicate feature strength. The ratio we derived for DI Cha was examined in comparison with other T Tauri and Herbig Ae/Be stars from the samples presented in \cite{2003A&A...412L..43P}, \cite{2006ApJ...639..275K}, and \cite{2009A&A...507..327O}, as well as with representative solar system comets. The result is shown in the Figure~\ref{fig:flux_ratio}. Compared to other T Tauri and Herbig Ae/Be stars, we found that DI Cha A has a relatively high 11.3/9.8~$\mu$m flux ratio and low strength of the silicate band, based on both MIDI and Spitzer measurements. This result shows that large particles of dust are already present in the outer disk component of the DI Cha A, thus even in the outer disk the dust component is not pristine interstellar dust. The conclusions we derived in this subsection are fully consistent with the choices of dust properties obtained for the best fitting radiative transfer model.


\subsection{Time variability}
\label{sec:variability}

In Sect.~\ref{sec:Results} we found that the optical brightness of DI~Cha~A was constant within a few hundredths of a magnitude (apart from a sinusoidal trend in the TESS data attributed to stellar rotation), on both daily and annual timescales. On the contrary, the mid-infrared flux exhibited significant variations on both timescales. 
Our modeling indicates that in the WISE bands both the stellar photosphere and the disk thermal emission contributes significantly to the emitted flux. Since the stellar component is constant, the mid-IR variability must be originated from the disk. In the lack of variable irradiation of the dust grains by the central star, geometrical changes of the dust distribution provide the most plausible explanations for the changing infrared thermal emission.

In the case of the cavity model, the radiative transfer modeling suggests that the measured WISE W1-band emission mainly originates from the stellar photosphere and the inner component (21\% and 78\%, respectively) and only in small proportion from the outer component. In the W2 band these numbers are 
17\% and 81\%, respectively. These results imply that the variability must be connected to some rearrangement of the inner ring structure. In the following we will speculate about three possible scenarios.

The first possibility would be that the vertical scale height of the inner ring changes with time. To interpret our WISE observations, one would need to assume that the inner ring  has been developing an increasingly larger scale height between 2014 and 2020. In the lack of external heating, and thus increased temperature and pressure, this puffing-up process cannot be explained by vertical expansion of the gas. It is more likely that some kind of hydrodynamical instability acts and modifies the scale height. 
A higher scale height increases the intercepted stellar light, and thus boosts the thermal emission of the disk. However, to achieve a real increase in the infrared flux it is a precondition that the inner ring is at least partially optically thick in the mid-infrared, that may not be satisfied everywhere.

The second possibility is to assume a disk model where the location of the inner edge of the inner ring can change with time 
\citep[see e.g. ][]{2001AJ....121.3160C}.
 Since the star is still accreting, an inward flow of material is possible, whose interaction with the stellar magnetic field would result in a changing inner disk radius as the accretion rate fluctuates. Dust closer to the star will be hotter and radiate with higher intensity.
 
The third scenario would be to assume the appearance of fresh dust around the star. It may be a dust cloud lifted above the disk surface by turbulent processes, or might even be a giant collision between already built-up solid bodies, planetesimals or planetary embryos, that would produce instantaneously a huge amount of fresh dust. The latter event would be similar to the ones hypothetized in the so-called extreme debris disks \citep{balog2009}, although in those systems the central star is already on the main sequence, and the disk is gas-poor. If the variability of the inner disk in DI Cha is indeed caused by a planetasimal collision, then the inner disk may be a transient phenomenon in this system.

\begin{figure}[!ht]
    \includegraphics[width=\columnwidth]{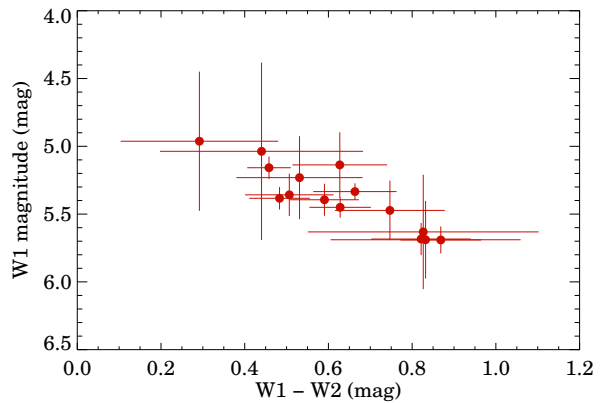} 
    \caption{Color-magnitude diagram of the WISE observations.}
    \label{fig:wisecolor}
\end{figure}

Theoretically, these scenarios could be distinguished by analyzing the color variations of the WISE light curves (Figure~\ref{fig:wisecolor}). The data suggest that the source becomes bluer when brighter, implying that the extra radiation arises from relatively hot dust particles. A higher precision mid-infrared monitoring of the object and a detailed modeling of the structural changes of the disk will be needed to further narrow down the most plausible physical mechanisms.

\subsection{Presence of planets}

\begin{figure}
    \centering
    \includegraphics[width=\columnwidth]{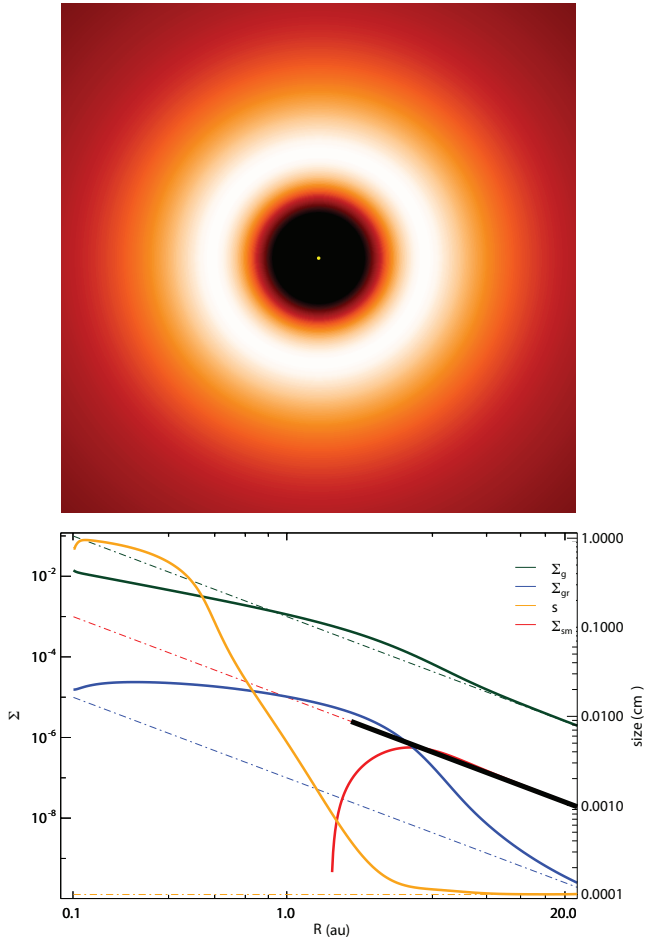}
    \caption{Snapshot of the hydrodynamic simulation that models the two component dust growth. \emph{Upper panel:} Surface mass density distribution of the small-sized dust component in the inner ($R<$8au) disk. \emph{Lower panel:} Radial profiles of gas (green) and the two dusty components (small-sized (red) and grown dust (blue)). Dot-dashed curves show the initial profiles. Orange line shows the representative particle size of grown dust. Note that the black lines represents the simple radiative transfer model of dusty component.}
    \label{fig:hydro_dust_growth}
\end{figure}

In our radiative transfer calculation we investigated two possible disk structures, the simple cavity and the gapped models. Here we discuss what processes might have resulted in these two disk geometries. 
 
 For the simple cavity model we evaluated
three possible scenarios. 
First, we performed calculations based on simple physics with a multiplanet system with 3 members. As a test, we estimated the possible planetary masses in the simple case where the 3 planets are in 1:2:4 orbital resonance.  The planetary Hill sphere is defined as

\begin{equation}
R_\mathrm{Hill} = R\left(\frac{q}{3}\right)^{\frac{1}{3}},
\end{equation}
 where $R$ is the distance from the central star, and $q$ is the planet-to-star mass ratio.
In this configuration, the planets cover the area from the sublimation radius of 0.18 au to a distance of 1.85 au from the star with their Hill-radii. The resulting mass is 707.32 M$_\mathrm{Jup}$ for the planet closest to the star, 86.01 M$_\mathrm{Jup}$ and 447.45 M$_\mathrm{Jup}$ for the other planets. The weakness of this scenario is the need for unrealistically large planetary masses.

For the next possible scenario, we performed 2D hydrodynamical simulations to model the interactions of protoplanetary disk gas and dust component with embedded planets. The details of the model are described in  \cite{2020MNRAS.497.5540R}. In the hydrodynamic model, the 3\,au hole is hypothetically carved by a single giant planet. Our simulations, however, revealed that the disk inside the planet is not depleted (via gas accretion) sufficiently to match the radiative transfer calculation. This is because both the gas and the well coupled dust component flow through the planetary orbit as it was first pointed out by \cite{2014ApJ...792L..10D}. 

To circumvent the issue of disk refilling, we prescribe planetary accretion by the method described by \cite{1999MNRAS.303..696K}. In this simple accretion method, the planet accretes the gas and dust from their Roche lobe by a certain amount of time. We found that a dust hole can be formed by assuming a relatively strong accretion, i.e. when the dust inside the Hill sphere is accreted within several orbits. Note that, the inner disk emptying was reproduced, but complete, i.e., an order of magnitude depletion can form in the inner disk.

We investigated a third possible scenario for cavity formation which is based on grain growth. We implemented the monodispersed grain growth model of \cite{1997A&A...319.1007S} based on \cite{2018A&A...614A..98V}. In this grain growth model, two dust components exist in the disk, the primordial small-sized dust and the grown dust. Due to the coagulation process, the small-sized dust grows and is transformed into the grown size component. Since the growth rate is inversely proportional to the stellar distance, the inner disk region starts to deplete in small-sized dust components forming an inner hole in micron-sized dust, see Fig.\ref{fig:hydro_dust_growth}.

The gapped disk model can be explained by the presence of a planetary system. However, explaining the presence of the inner dusty ring is not straightforward. According to the theory, an accretionary inactive dead zone might exist in the disk \citep{1996ApJ...457..355G}, in which case a pressure maximum can form at the dead zone edges due to accretion mismatch. This pressure maximum is capable to collect dust particles to form an inner dusty ring at the inner edge of the dead zone. However, the dust clearing is hindered due to the strongly reduced viscosity (at least an order of magnitude) in the dead zone where the gap exists. The complexity of the theory is further increased by the fact that even larger planets may be responsible for creating the larger extent gap. We can assume that this problem can be solved if there are more than three planets, but this makes it even more difficult to perform a specific study. 

Here we have to emphasize that the grain growth model is not feasible in the case of the gapped disk model. Since the grain growth is presumably very effective at the pressure maximum at the inner dead zone edge, the inner dusty ring is depleted with micron-sized dust. This would mean precisely that we cannot detect as much dust development in the inner dead zone edge as in the areas a few au away, which contradicts the experience so far.






\section{Summary}

We performed a detailed interferometric study of the brightest star of the DI Cha quadruple system. In addition to VLTI/MIDI and VLTI/PIONIER interferometric measurements, we used several other photometric measurements and light curves in our work. We revealed the structure of the inner disc region by geometric and radiative transfer model fitting. The best fit was achieved with a model with a hot inner disk close to a star between 0.19 and 0.21 au. A gap was present in the disk between 0.21 and 3.0 au,  which is a characteristic phenomenon  in the geometry of pre-transitional disks. From 3 au, a classical disc component was included, which had a puffed up inner rim. In our model the inner disk contained more large-sized dust than the outer disk due to the sloping dust particle size distribution.  This confirms the assumption that more progressed dust is present in the innermost areas.
In our detailed light curve analysis, we found that although the radiation of the star is constant, the inner disk shows variability in the mid-infrared wavelengths, presumably due to geometric changes of the inner disk. We also examined the reasons for the formation of the discovered geometry and described some possible planet forming scenarios.

\section*{Acknowledgments}
The project was supported by the Hungarian OTKA grant K132406 and the \'UNKP-20-3 New National Excellence Program of the Ministry for Innovation and Technology from the source of National Research, Development and Innovation fund.
It has also received funding from the European Research Council (ERC) under the European Union's Horizon 2020 research and innovation programme under grant agreement No 716155 (SACCRED).
ZsR was supported by the Hungarian OTKA Grant No. 119993.
This work has made use of data from the European Space Agency (ESA) mission {\it Gaia} (\url{https://www.cosmos.esa.int/gaia}), processed by the {\it Gaia} Data Processing and Analysis Consortium (DPAC, \url{https://www.cosmos.esa.int/web/gaia/dpac/consortium}). Funding for the DPAC has been provided by national institutions, in particular the institutions participating in the {\it Gaia} Multilateral Agreement.
The Combined Atlas of Sources with Spitzer IRS Spectra (CASSIS) is a product of the IRS instrument team, supported by NASA and JPL. CASSIS is supported by the "Programme National de Physique Stellaire" (PNPS) of CNRS/INSU co-funded by CEA and CNES and through the "Programme National Physique et Chimie du Milieu Interstellaire" (PCMI) of CNRS/INSU with INC/INP co-funded by CEA and CNES.
This research has made use of the Jean-Marie Mariotti Center \texttt{OiDB} service \footnote{Available at http://oidb.jmmc.fr }.
%

\vspace{5mm}
\facilities{VLTI(MIDI), VLTI(PIONIER), WISE, {\it Gaia}, ASAS, TESS}

\bibliography{ms}{}
\bibliographystyle{aasjournal}

\end{document}